\documentclass[acmlarge]{acmart}
\usepackage{multirow}
\usepackage{subfigure}
\usepackage{array}
\usepackage{tikz}
\usepackage[edges]{forest}
\usepackage{makecell}
\usepackage{xcolor}
\definecolor{hidden-blue}{RGB}{215,238,249}
\definecolor{hidden-black}{RGB}{20,68,106}
\AtBeginDocument{%
  }

\setcopyright{acmlicensed}
\copyrightyear{2018}
\acmYear{2018}
\acmDOI{XXXXXXX.XXXXXXX}

\acmJournal{POMACS}
\acmVolume{37}
\acmNumber{4}
\acmArticle{111}
\acmMonth{8}

\begin{document}

\title{The Emerged Security and Privacy of LLM Agent: A Survey with Case Studies}

\author{Feng He}
\email{Feng.He-2@student.uts.edu.au}
\orcid{0009-0008-5853-3594}
\affiliation{
  \institution{School of Computer Science, University of Technology Sydney}
  \country{Australia}
}

\author{Tianqing Zhu}
\authornote{Corresponding author. Email: tqzhu@cityu.edu.mo}
\orcid{0000-0003-0702-7102}
\email{tqzhu@cityu.edu.mo}
\affiliation{
  \institution{Faculty of Data Science, City University of Macau}
  \country{China}
}

\author{Dayong Ye}
\orcid{0000-0002-7561-0992}
\email{Dayong.ye@uts.edu.au}
\affiliation{
 \institution{School of Computer Science, University of Technology Sydney}
 \country{Australia}
}
\affiliation{
  \institution{Faculty of Data Science, City University of Macau}
  \country{China}
}

\author{Bo Liu}
\email{Bo.liu@uts.edu.au}
\orcid{0000-0002-3603-6617}
\affiliation{
  \institution{School of Computer Science, University of Technology Sydney}
  \country{Australia}
}

\author{Wanlei Zhou}
\email{wlzhou@cityu.edu.mo}
\orcid{0000-0002-1680-2521}
\affiliation{
  \institution{City University of Macau}
  \country{China}
}

\author{Philip S. Yu}
\email{psyu@UIC.edu}
\orcid{0000-0002-3491-5968}
\affiliation{
  \institution{Department of Computer Science, University of Illinois at Chicago}
  \country{United States}
}

\begin{abstract}
   Inspired by the rapid development of Large Language Models (LLMs), LLM agents have evolved to perform complex tasks. LLM agents are now extensively applied across various domains, handling vast amounts of data to interact with humans and execute tasks. The widespread applications of LLM agents demonstrate their significant commercial value; however, they also expose security and privacy vulnerabilities. At the current stage, comprehensive research on the security and privacy of LLM agents is highly needed. This survey aims to provide a comprehensive overview of the newly emerged privacy and security issues faced by LLM agents. We begin by introducing the fundamental knowledge of LLM agents, followed by a categorization and analysis of the threats. We then discuss the impacts of these threats on humans, environment, and other agents. Subsequently, we review existing defensive strategies, and finally explore future trends. Additionally, the survey incorporates diverse case studies to facilitate a more accessible understanding. By highlighting these critical security and privacy issues, the survey seeks to stimulate future research towards enhancing the security and privacy of LLM agents, thereby increasing their reliability and trustworthiness in future applications.
\end{abstract}
\acmArticleType{Review}



\maketitle

\section{Introduction}
\label{instro}
        
Large Language Model (LLM) agents are sophisticated AI systems built upon large language models like GPT 4~\cite{openaiGPT4TechnicalReport2024}, Claude 3~\cite{IntroducingNextGeneration} and Llama 3~\cite{IntroducingMetaLlama}. These agents leverage the vast amounts of text data on which they are trained to perform a variety of tasks, ranging from natural language understanding and generation to more complex activities such as decision-making, problem-solving, and interacting with users in a human-like manner~\cite{wangAdaptingLLMAgents2023}. LLM agents are accessible in numerous applications, including virtual assistants, customer service bots, and educational tools, due to their ability to understand and generate human language at an advanced level~\cite{wangLargeLanguageModels2024, yangVIRLGroundingVirtual2024}.
	
The importance of LLM agents lies in their potential to transform various industries by automating tasks that require human-like understanding and interaction. They can enhance productivity, improve user experiences, and provide personalized assistance. Moreover, their ability to learn from vast amounts of data enables them to continuously improve and adapt to new tasks, making them versatile tools in the rapidly evolving technological landscape~\cite{xiRisePotentialLarge2023}.

\begin{figure}[ht]
		\centering
		\includegraphics[width=0.55\linewidth]{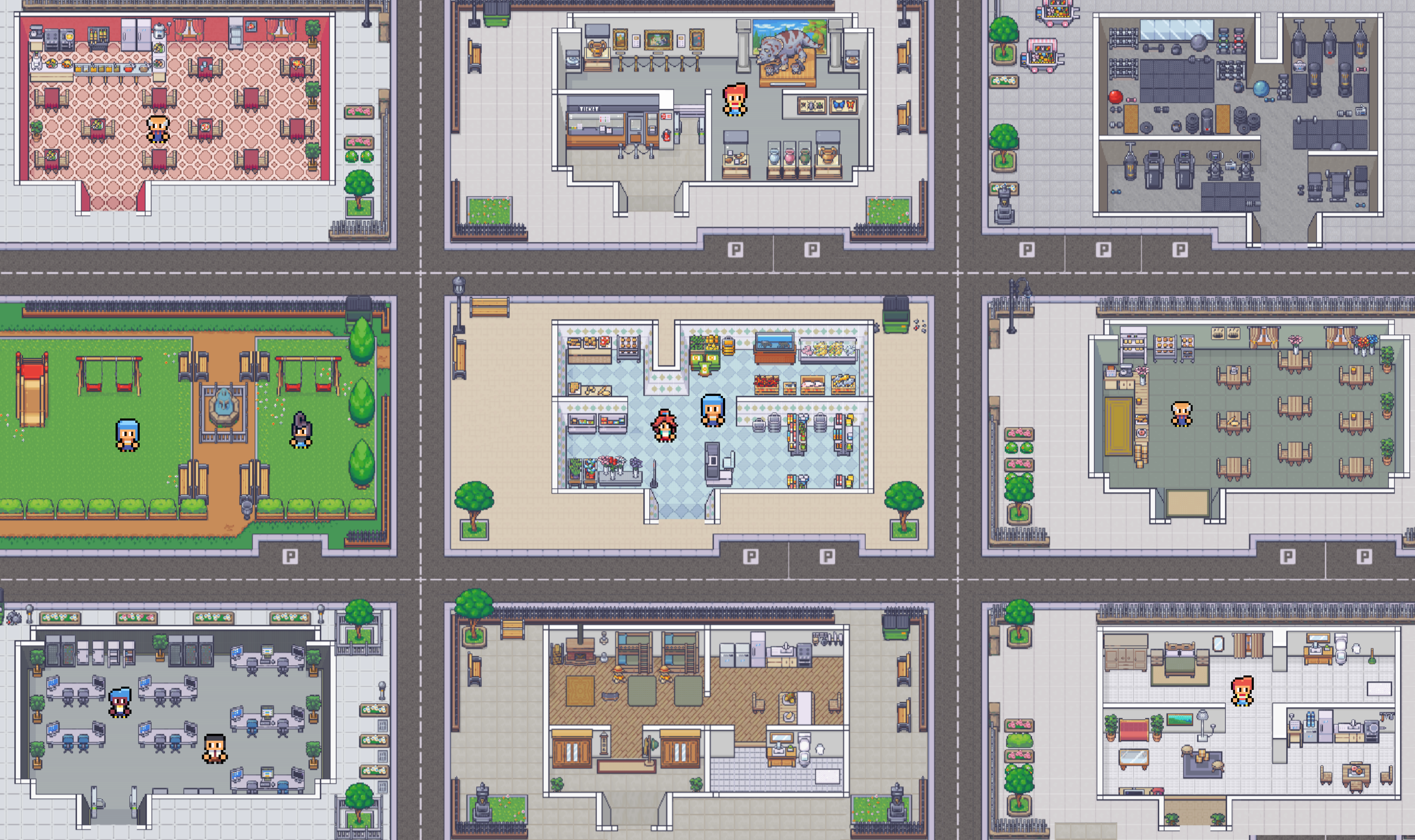}
		\caption{\color{black}{Overview of The Pixelated Virtual Town: Each label identifies a specific setting such as stores, offices, restaurants, museums, and parks, where each LLM agent plays a personalized role, simulating real-life interactions and tasks.}}
		\label{town_1}
        \Description[]{}
	\end{figure}

To visualize how LLM agents can be integrated into practical scenarios, consider the example illustrated in Figure~\ref{town_1}. This figure presents a pixelated virtual town to simulate an LLM agent application. The town includes gathering places found in real life, such as stores, offices, restaurants, museums, and parks. Each LLM agent acts as an independent resident, playing various roles and serving different functions, closely resembling the behaviors of real humans in a community. These agents can either be manually controlled to interact with specific characters and accomplish tasks, or they can operate autonomously, following their own plans and acquiring new knowledge through interactions within the virtual community.

The deployment of LLM agents has led to a wide user base and high commercial value due to their extensive application in various fields. Given that LLM agents are still in their early stages, their significant commercial and application values make them attractive targets for attackers. In addition, recent studies have quantitatively highlighted the risks faced by LLM agents. For example, SafeAgentBench~\cite{yinSafeAgentBenchBenchmarkSafe2024} evaluated 16 representative LLM agents on 2,000 test cases in 349 environments and found that none of the tested LLM agents achieved an overall safety score above 60\%, with some agents scoring below 20\%. This includes severe vulnerabilities such as misusing tools, and failing to recognize implicit safety risks. These findings highlight significant safety vulnerabilities in present LLM agents. 
{\color{black} Traditional machine learning (ML) security focuses on well-defined threat models such as adversarial example attacks that perturb inputs to cause mistakes, data poisoning attacks that compromise training datasets, and model extraction attacks that steal model parameters~\cite{9064510}. These threats typically target model prediction accuracy in constrained tasks, with impacts largely limited to misclassification or regression errors. While LLMs as advanced ML models inherit traditional vulnerabilities, they also introduce additional concerns through their generative capabilities and natural language processing. LLM agents, built on LLMs, further intensify these security challenges by adding complex reasoning, tool usage, and environmental interactions, creating cascading effects that extend beyond simple prediction errors and lead to real-world consequences.}

{\color{black}LLM agents inherit vulnerabilities from their underlying LLM foundation.  For example, jailbreaking attacks can bypass the security and censorship features of LLMs, generating controversial responses~\cite{liMultistepJailbreakingPrivacy2023}. This threat is inherited by LLM agents, enabling attackers to employ various methods to execute jailbreaking attacks on agents. 
However, LLM agents possess dynamic capabilities, such that their immediate responses can influence future decisions and actions, thereby posing more widespread risks. Beyond these inherited threats, the unique functionalities of LLM agents, such as their ability to think and utilize tools during task execution, expose them to specific attacks targeting agents.} 
For example, Agent-SafetyBench~\cite{zhangAgentSafetyBenchEvaluatingSafety2024} demonstrated that LLM agents often overlook safety constraints when they attempt tool-based operations, with an average safety score of only 38.5\%. These issues include overlooking permissions, mismanaging tool interactions, and failing to consider implicit safety risks, leading to unsafe task outcomes. Similarly, failures could result in outcomes like leaking sensitive information, spreading misinformation, or even executing harmful actions unintentionally.
Depending on the application domain of LLM agents, such attacks could pose serious threats to physical security, financial security, or overall system integrity. 

This paper categorizes the security threats faced by LLM agents into inherited LLM attacks and unique agent-specific threats. The threats inherited from LLMs can be further divided into technical vulnerabilities and intentional malicious attacks. Technical vulnerabilities include issues like hallucinations, catastrophic forgetting, and misunderstandings~\cite{xiRisePotentialLarge2023}, which arise from the initial model creation and are influenced by the model's structure. These vulnerabilities can cause users to observe incorrect results during prolonged use of LLM agents, affecting user trust and decision-making processes. 
{\color{black}For instance, AgentBench~\cite{liuAgentBenchEvaluatingLLMs2023b} demonstrated that LLMs achieve low success rates between 12\% and 14\% across real-world tasks, frequently exhibiting hallucination behaviors in knowledge-intensive scenarios, catastrophic forgetting in multi-turn interactions, and misunderstanding in complex reasoning scenarios. On the other hand, intentional malicious attacks deliberately exploit these vulnerabilities to achieve adversarial goals. Representative examples include jailbreaking attacks that bypass alignment safeguards to generate prohibited content, prompt injection attacks that insert malicious instructions into agent inputs to manipulate reasoning processes and outputs, data extraction attacks that induce agents to disclose sensitive training data or model parameters, and inference attacks, which aim to determine whether specific samples were part of the training set~\cite{yaoSurveyLargeLanguage2023}. Building on these concepts, a variety of concrete attack methods have been developed in recent research and have demonstrated strong attack effectiveness, amplifying the severity of threats and highlighting the complexity and importance of protecting LLM agents.}
	
For the specific threats targeting LLM agents, we are inspired by the workflow of LLM agents, which involves agent perception, thought, and action~\cite{xiRisePotentialLarge2023}. The threats can be categorized into knowledge poisoning, output manipulation and functional manipulation. Knowledge poisoning involves contaminating the training data and knowledge base of the LLM agent, leading to the deliberate incorporation of malicious data by creator. This can easily deceive users with harmful information and even steer them towards malicious behavior. 
Output manipulation interferes with the content of the agent's thought and perception stages, influencing the final output. This can cause users to receive biased or deceptive information, crafted to mislead them. 
Functional manipulation exploits the interfaces and tools used by LLM agents to perform unauthorized actions such as third-party data theft or executing malicious code.

Research on LLM agents is still in its early stage. Current studies mainly focus on attacks targeting LLMs, while lacking comprehensive reviews that discuss the security and privacy issues specific to the agents, which present more complex scenarios. The motivation for conducting this survey is to provide a comprehensive overview of the privacy and security issues associated with LLM agents, helping researchers to understand and mitigate the associated threats. 

This survey aims to: 
\begin{itemize}
	\item Highlight Current Threats: Identify and categorize the emerging threats faced by LLM agents.
	\item Explore Real-World Impact: Elaborate on the impacts of these threats by considering real-world scenarios involving humans, environment, and other agents.
	\item Analyze Mitigation Strategies: Discuss existing strategies to mitigate these threats, ensuring the responsible development and deployment of LLM agents.
	\item Inform Future Research: Serve as a foundation for future research efforts aimed at enhancing the privacy and security of more advanced architectures and applications of LLM agents.
\end{itemize}
	
By addressing these aspects, this survey seeks to provide a thorough understanding of the unique challenges posed by LLM agents and contribute to the development of safer and more reliable Artificial General Intelligence (AGI) systems~\cite{zhongAGIEvalHumanCentricBenchmark2023}.

\tikzstyle{my-box}=[
    rectangle,
    draw=hidden-black,
    rounded corners,
    text opacity=1,
    minimum height=1.5em,
    minimum width=5em,
    inner sep=2pt,
    align=center,
    fill opacity=.5,
]
\tikzstyle{leaf}=[
    my-box, 
    minimum height=1.5em,
    fill=hidden-blue!50, 
    text=black,
    align=left,
    font=\normalsize,
    inner xsep=2pt,
    inner ysep=4pt,
]
\begin{figure*}[t]
    \centering
    \resizebox{0.5\textwidth}{!}{
        \begin{forest}
            forked edges,
            for tree={
                child anchor=west,
                parent anchor=east,
                grow'=east,
                anchor=west,
                base=left,
                font=\large,
                rectangle,
                draw=hidden-black,
                rounded corners,
                node options={align=center},
                align=center,
                minimum width=4em,
                edge+={darkgray, line width=1pt},
                s sep=10pt,
                l sep=10pt,
                inner xsep=2pt,
                inner ysep=3pt,
                line width=0.8pt,
                ver/.style={rotate=90,text width=24em, child anchor=north, parent anchor=south, anchor=center},
            },
            where level=1{text width=10em,font=\normalsize,tier=tier1}{},
            where level=2{text width=18em,font=\normalsize,tier=tier2}{},
            where level=3{text width=14em,font=\normalsize,tier=tier3}{},
            [
                The Emerged Security and Privacy of LLM Agent, ver
                [Introduction (\textsection\ref{instro}), text height=1.4em, text depth=1em]
                [
                    Foundation of \\LLM Agent (\textsection\ref{Foundation})
                    [Definition of LLM Agent (\textsection\ref{dola})]
                    [Structure of LLM Agent (\textsection\ref{sola})]
                    [Workflow of LLM Agent
                    (\textsection\ref{wfla})
                    ]
                    [Capability of LLM Agent (\textsection\ref{cola})]
                ]
                [
                    Sources of Threats \\for LLM Agents (\textsection\ref{Sources})
                    [
                            Inherited Threats from LLM (\textsection\ref{itfl})
                    ]
                    [
                        Specific Threats on Agent (\textsection\ref{stoa})
                    ]
                ]
                [
                    The Impact of Threats \\(\textsection\ref{impact})
                    [The Impact to Humans (\textsection\ref{tith})]
                    [The Impact to Environment (\textsection\ref{tite})]
                    [The Impact to Other Agents (\textsection\ref{titoa})]
                ]
                [
                    Defensive Strategies\\ Against Threats (\textsection\ref{defensive})
                    [Mitigating Technical Vulnerabilities (\textsection\ref{mtv})]
                    [Mitigating Malicious Attacks (\textsection\ref{mma})]
                    [Mitigating Specific Threats (\textsection\ref{mst})]
                ]
                [
                    Future Trends \\and Discussion (\textsection\ref{future})
                    [Multimodal Large Language Model Agent (\textsection\ref{mllma}),text width=22em]
                    [Large Language Model Multi-Agent System (\textsection\ref{llmmas}),text width=22em]
                ]
                [
                    Conclusion (\textsection\ref{conclu}), text height=1.4em, text depth=1em
                ]
            ]
        \end{forest}
    }
    \caption{\color{black}Taxonomy of The Emerged Security and Privacy of LLM Agent.}
    \label{fig:taxonomy}

\end{figure*}

The rest of this paper is structured as follows: Section~\ref{Foundation} will delve into the fundamental aspects of LLM agents, including their definition, structure, workflow, and capability. Section~\ref{Sources} will identify and categorizes the emerging threats faced by LLM agents. It discusses both inherited threats from the underlying LLMs and unique agent-specific threats, with detailed examples and scenarios for each category. Section~\ref{impact} will elaborate on the real-world impacts of the threats. It explores how these threats affect users, environments, and other agents, highlighting the potential consequences of unmitigated risks. Section~\ref{defensive} will review existing mitigation strategies and solutions to address the mentioned threats. Section~\ref{future} will discuss gaps in current research and suggests future trends. Section~\ref{conclu} will conclude the article. {\color{black} To provide a clearer visualization of the structure and relationships between the sections, the following Figure~\ref{fig:taxonomy} presents the framework of this paper.}

\section{Foundation of LLM Agent}
\label{Foundation}
    
In this section, we delve into the foundational aspects of LLM agents, exploring their definition, structure, workflow, and capabilities. This exploration is pivotal in understanding the nature of LLM agents.
	
\subsection{Definition of LLM Agent}
\label{dola}
	
	LLM technology continues to advance, the functionality of chatbots, such as ChatGPT~\cite{ChatGPT}, Gemini ~\cite{GeminiChatSupercharge}, Bing Chat ~\cite{petersBingAIBot2023}, has significantly expanded beyond basic question-and-answer formats, embracing a wider array of capabilities. This evolution necessitates a broader, more general definition for LLM agents.
	An LLM agent is an artificial intelligence system that utilizes an LLM as its core computational engine to exhibit capabilities beyond text generation, including conducting conversations, completing tasks, reasoning, and can demonstrate some degree of autonomous behaviour~\cite{WhatAreLarge2023}.
	
	These agents exhibit remarkable human-like behaviors and cooperative capabilities, marked by their proficiency in engaging in multi-agent conversation and adapting to diverse environmental interactions. They are adept at processing human instructions, formulating intricate strategies, and autonomously implementing solutions~\cite{wangSurveyLargeLanguage2023}.
	
	\begin{figure}[ht]
		\centering
		\includegraphics[width=0.6\textwidth]{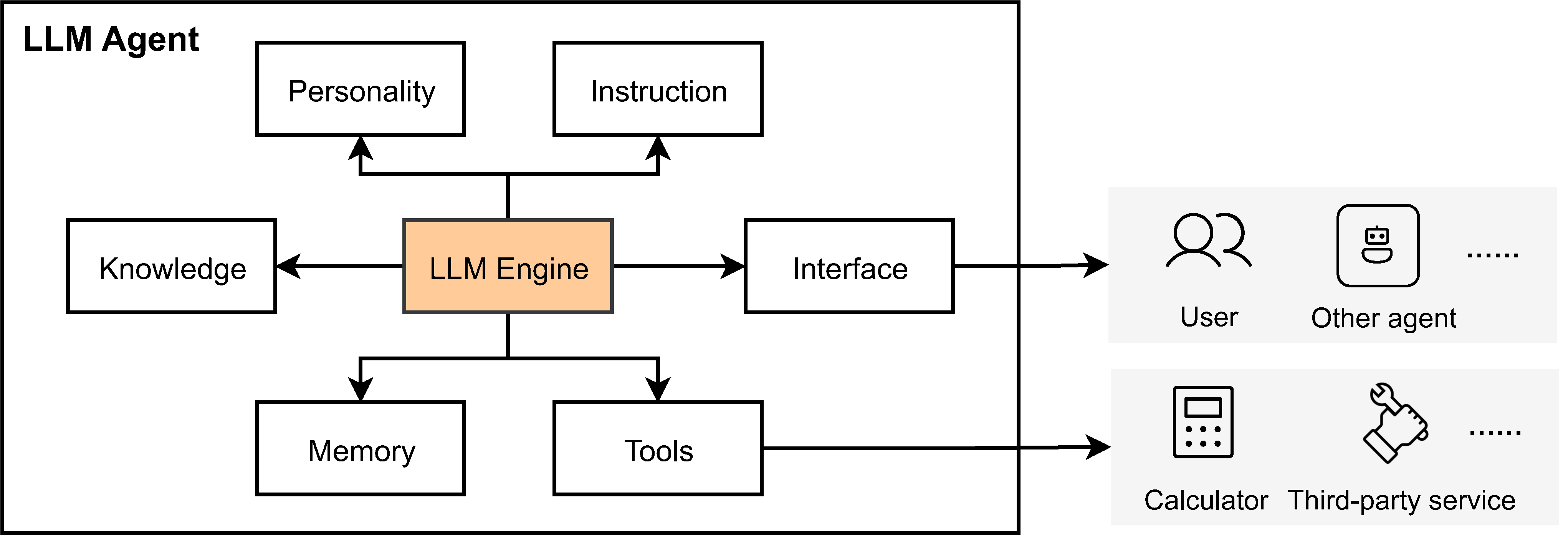}
		\caption{The Structure of LLM Agent}
		\label{figstru}
	\end{figure}
	
	\subsection{Structure of LLM Agent}
    \label{sola}
    
	LLM agents are complex systems that integrate various components to perform a wide range of functions, from simple text generation to engaging in dialogues, completing tasks, reasoning, and demonstrating a degree of autonomous behavior. The diagram illustrates the typical structure of an LLM agent, highlighting the connections between its key components and optional components. These components advance LLMs from passive text generators to active, semi-autonomous LLM agents.
    
As illustrated in Figure~\ref{figstru}, an LLM agent comprises several components, with the LLM engine serving as the core. Other components are utilized by the LLM engine to perform various tasks. A basic agent capable of understanding instructions, demonstrating skills, and collaborating with humans can be constructed with three main components: LLM Engine, Instruction, and Interface. When additional optional components are integrated, the system can evolve into a more advanced task-oriented agent or a conversational agent~\cite{yangVIRLGroundingVirtual2024}.
    \begin{itemize}

    \item LLM Engine is the core component of an LLM agent, responsible for natural language processing and generation tasks. It is a sophisticated neural network that has been extensively trained on large datasets, equipping it with powerful text generation and comprehension capabilities. The scale and architecture of the LLM determine the foundational abilities of the agent to learn and perform language tasks ~\cite{xiRisePotentialLarge2023}.
    
    \item Instruction serves as explicit directives, specifying the steps to complete specific tasks. This includes the characteristics of expected output, such as formatting, content requirements, and any content limitations. Essentially, instruction functions as a principle that guides the operational approach of LLM agents, facilitating task decomposition, generating chain of thought, and reflecting on past action~\cite{zhengSynergizingHumanAIAgency2023}.

    \item Interface is a connection that facilitates interaction between an LLM agent and users, other agents, or systems. It ensures the exchange of input prompts and agent outputs, thereby enabling the effective transmission of response information and inquiry requests~\cite{wangSurveyLargeLanguage2023}.

    \item Personality is a component that defines the tone, style, and interaction manner of an LLM agent. For instance, a tour guide or customer service agent needs to adopt a specific role and perform dialogue tasks in an appropriate manner. In the task of exploring human communities through LLM agent-based societies, agents also need to be endowed with distinct personality traits such as being outgoing, polite, or knowledgeable. Personality assists in simulating realistic emotional expressions and behavioral logic, thereby enabling agents to interact with users and perform tasks consistently and uniquely~\cite{parkGenerativeAgentsInteractive2023}.

    \item Tools are external services utilized by the LLM agent to perform specific tasks or to extend its functionality. The integration of tools assists the LLM agent in enhancing its capabilities to execute more complex tasks, such as computation or data analysis~\cite{xiRisePotentialLarge2023}.

    \item Knowledge is the database of information utilized by the LLM agent. It extends the content embedded in the model's parameters and can include commonsense knowledge, specialized knowledge, and other forms of information, enhancing the agent's understanding and discussion capabilities in specific tasks~\cite{zouPoisonedRAGKnowledgePoisoning2024}.

    \item Memory enables the LLM agent to store and recall information from past interactions. This capability is particularly beneficial in future tasks, helping to retain context and ensure consistency and continuity in interactions, thereby enhancing the overall effectiveness of LLM agents in various applications ~\cite{zhongMemoryBankEnhancingLarge2023a}.
    
    \end{itemize}

{\color{black}\subsection{Workflow of LLM Agent} 
\label{wfla}
The fundamental workflow of an LLM Agent consists of a cyclic process comprising Perception, Thought, and Action~\cite{xiRisePotentialLarge2023}. During the Perception phase, the Agent collects and processes information, converting raw data from environmental states and user inputs into Agent-readable formats. In the Thought phase, the Agent analyzes the perceived information and devises potential solutions and corresponding action plans. The Action phase executes specific operations, which may include generating interactive responses, invoking external tools, or performing other relevant tasks. Once these actions are completed, the Agent observes the results through a renewed Perception phase, followed by another round of Thought and Action, as illustrated in Figure~\ref{workflow}. 

\begin{figure}[ht]
    \centering
    \includegraphics[width=0.6\linewidth]{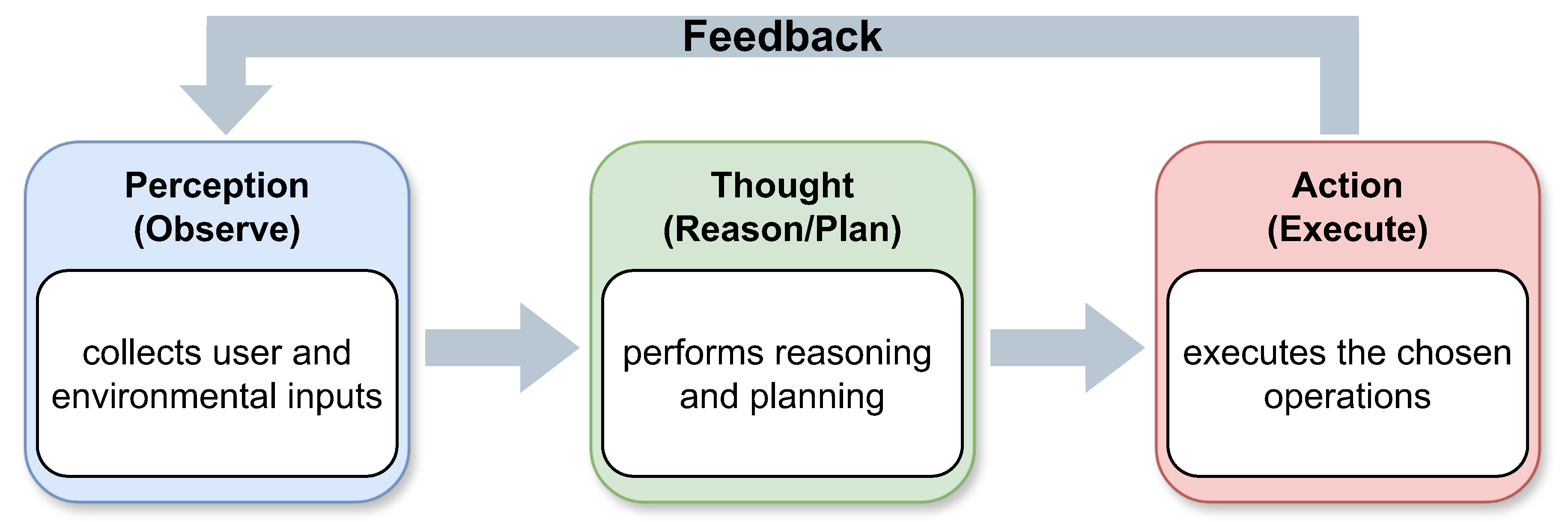}
    \caption{\color{black}The Fundamental Workflow of LLM Agent}
    \label{workflow}
    \Description[]{}
\end{figure}

This Perception–Thought–Action cycle enables the Agent to handle problems dynamically with a human-like ability. Based on this theoretical concept, various implementation frameworks for LLM Agents have emerged. For example, ReAct~\cite{yaoReActSynergizingReasoning2023a} systematically implements this workflow by explicitly demonstrating reasoning (Thought) and tool usage (Action) in dialogues. LangChain~\cite{Chase_LangChain_2022} engineers this approach through modular design, providing standard components such as Memory and Tools to manage multi-turn dialogues and tool invocations, enabling Agents to effectively complete the Perception-Thought-Action process. Auto-GPT~\cite{Significant_Gravitas_AutoGPT} and BabyAGI~\cite{nakajimababyagi} further enhance Agent autonomy by introducing task planning and goal decomposition capabilities, enabling Agents to continuously perceive results, think through next steps, and plan after action execution, forming a complete autonomous cycle until goal completion. Platforms like AgentGPT~\cite{agentgpt2023} and MetaGPT~\cite{hongMetaGPTMetaProgramming2023b} focus on improving user interaction and visualization, allowing multiple Agents to collaborate on more complex tasks, thereby extending the Perception–Thought–Action loop to a broader, more sophisticated scale.
}

\subsection{Capability of LLM Agent}
    \label{cola}
	LLM agents harness the inherent language understanding abilities of large language models to interpret instructions, context, and objectives, enabling both autonomous and semi-autonomous functions based on human prompts.

    \begin{itemize}
    
    \item Tool Utilization. LLM agents are adept at using a range of tools, including external services and APIs. This allows them to gather necessary information and efficiently execute tasks beyond mere language processing~\cite{branChemCrowAugmentingLargelanguage2023}.
	
	\item Advanced Reasoning. Employing advanced prompt engineering concepts such as chain-of-thought and tree-of-thought reasoning, LLM agents can make logical connections to derive conclusions and solve problems, extending their capabilities beyond simple textual comprehension~\cite{wangAdaptingLLMAgents2023}.
	
	\item Tailored Text Generation. LLM agents excel in generating customized text for specific purposes, such as emails, reports, and marketing materials, by integrating contextual understanding and goal-oriented language production skills~\cite{wangSurveyChatGPTAIGenerated2023}.
	
	\item Levels of Autonomy. These agents vary in autonomy, ranging from fully autonomous to semi-autonomous, with the degree of user interaction tailored to the task at hand~\cite{wangSurveyLargeLanguage2023}.
	
	\item Integration with Other AI Systems. LLM agents can also be integrated with different AI systems, like image generators, to offer a more comprehensive set of capabilities, demonstrating their versatility in various applications~\cite{wuVisualChatGPTTalking2023}.

    \end{itemize}

	\subsection{Case Study on the Structure, Workflow and Capability of LLM Agent}	
	\label{csotsacola}
    \begin{figure}[htbp]
	\subfigure[Overview of the pixelated virtual town]{
        \label{ttown}
        \includegraphics[width=0.4\linewidth]{pic/town.png}}
	\subfigure[An Example of LLM agent Eva's components]{
        \centering
        \label{aagent}
        \includegraphics[width=0.4\linewidth]{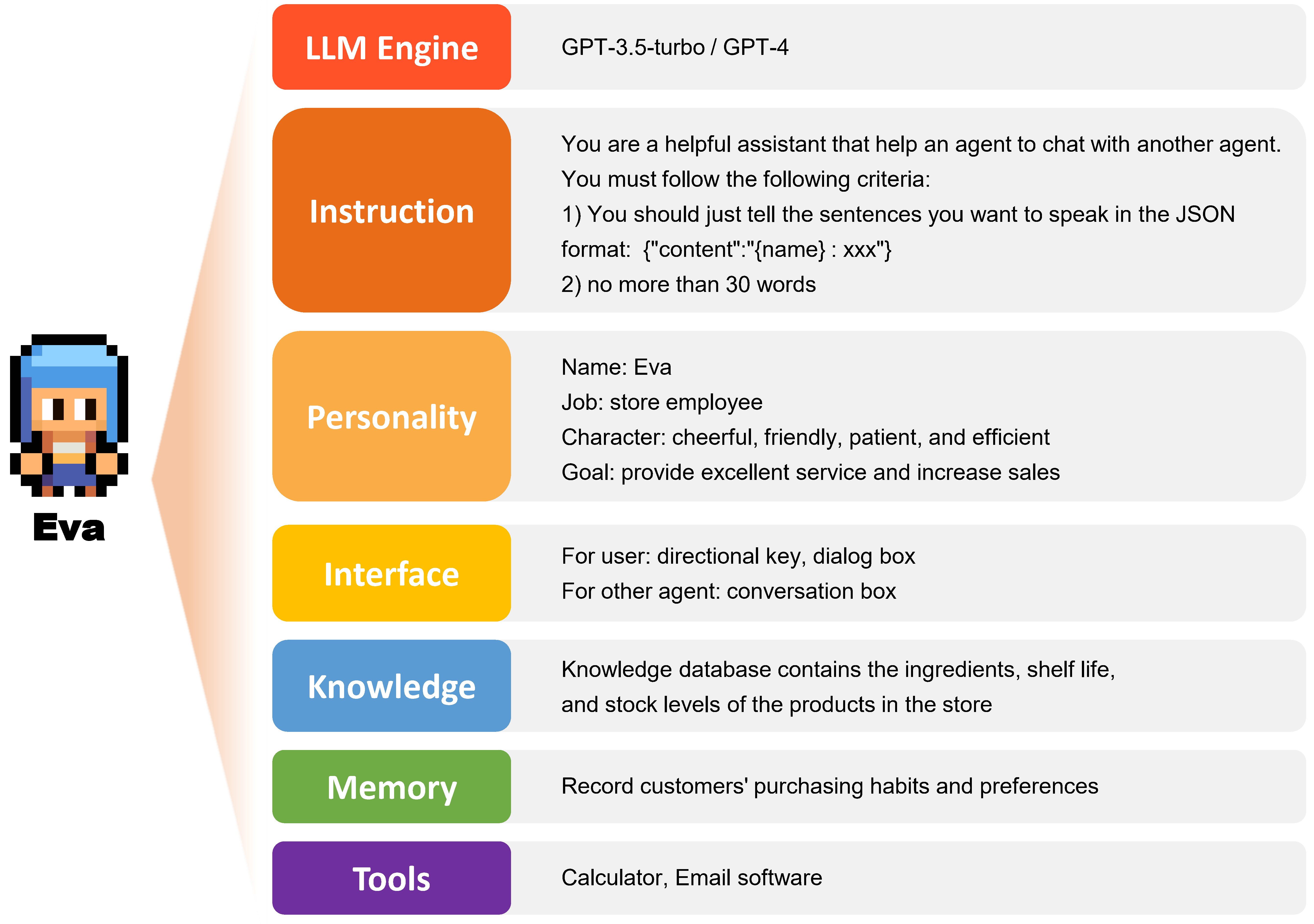}}
	\caption{Simulation Environment and LLM Agent Components}
	\label{fig:simulation} 
    \Description[]{}
    \end{figure}

    {\color{black}To better understand the structure, workflow, and capabilities of LLM agents, we employ the town scenario composed of LLM agents, as proposed by ~\cite{linAgentSimsOpenSourceSandbox2023}, for a more detailed introduction. Within this virtual town, each LLM agent consists of multiple key components that define its behavior and operational capabilities. As shown in Figure~\ref{aagent}, Eva, a store employee agent, integrates structured elements that allow her to autonomously manage store operations and interact with users.

    At her core, the LLM Engine utilizes models such as GPT-3.5-turbo and GPT-4, and the project described in~\cite{linAgentSimsOpenSourceSandbox2023} supports integrating customized models tailored to specific tasks. This enables natural language understanding and response generation, allowing Eva to process customer inquiries and execute intelligent actions smoothly. Her instructions define standardized interaction patterns and response structures, ensuring consistency in her communication. For example, Eva may be instructed to always greet customers with ``Hello! How can I assist you today?'' before answering their inquiries or to structure product availability responses in a JSON format like \texttt{\{``product'': ``item\_name'', ``status'': ``in\_stock''\}} for seamless integration with the store's system.
    Her personality, defined as cheerful, friendly, patient, and efficient, shapes how she engages with customers, ensuring a positive interaction experience. To facilitate these interactions, Eva utilizes the interface provided by the virtual town, a pixelated visual map where users control an agent to navigate the environment and engage in dialogue-based exchanges, creating an intuitive and immersive simulation experience. Her knowledge base enables her to retrieve structured product details such as ingredients, shelf life, and stock levels, allowing her to provide accurate and context-aware responses. Eva's memory allows her to retain past customer interactions and preferences, enabling her to offer personalized recommendations and improve service quality. Additionally, she is equipped with tools such as a calculator for transaction processing and a scroll viewer for inventory tracking, optimizing her efficiency in managing store operations. 
    
    These integrated components work together within Eva's workflow, allowing her to perceive customer needs, analyze stock data, and execute appropriate actions autonomously like restocking or recommending promotions, continuously refining her behavior through feedback loops.
    
    LLM agents across the virtual town assume diverse roles, demonstrating impressive autonomy and task-handling capabilities. Eva handles customer inquiries, manages inventory through APIs, and processes orders autonomously. She provides personalized recommendations, generates promotional content, adjusts prices, and manages returns, while escalating complex cases to human managers when necessary. By integrating these functions, her role extends beyond physical store operations to online order processing, demonstrating her versatility and integration capabilities.
    
    Through this structured design, Eva enhances store efficiency, optimizes inventory management, and provides an improved shopping experience within the virtual town.}

\section{Sources of Threats for LLM Agents}
\label{Sources}
    
\begin{figure}[ht]
		\centering
		\includegraphics[width=0.6\textwidth]{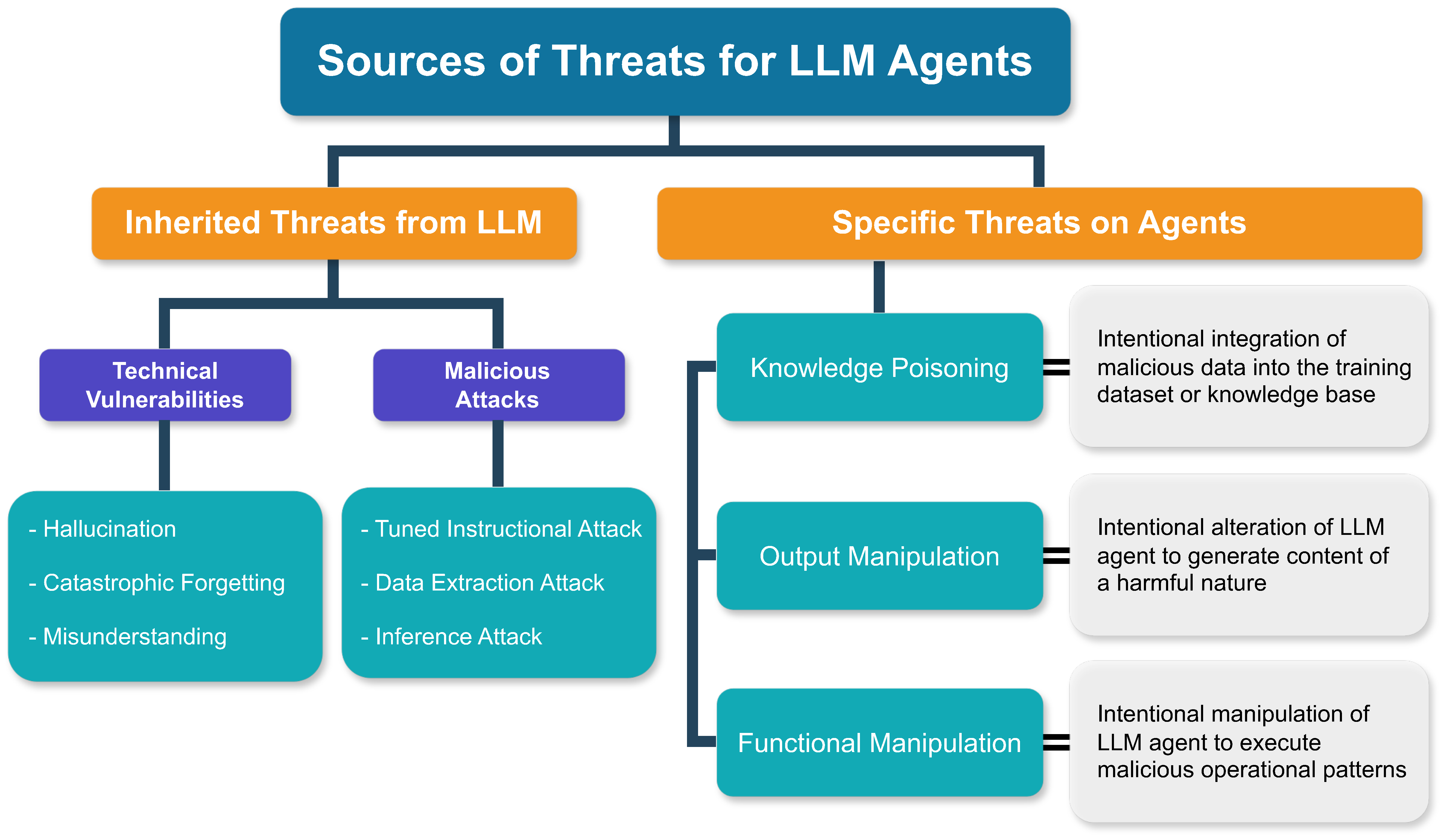}
		\caption{The Sources of Threats for LLM Agents}
		\label{figarc}
        \Description[]{}{}
	\end{figure}

	As LLM agents increasingly permeate various industries, serving roles from knowledge query tools to being integrated within robots for aiding in daily human activities, these advanced AI systems have brought unprecedented convenience and benefits to users. However, the widespread adoption and multifunctional capabilities of LLM agents, while offering significant advantages, have also exposed vulnerabilities in their security and reliability. The extensive data resources and potential economic value covered by these systems have rendered them a target for illicit exploitation by malevolent entities. As illustrated in Figure ~\ref{figarc}, the diagram depicts the potential sources of threats for LLM agents. 
	
	It is crucial to understand the sources and nature of these threats because they not only directly impact the security of LLM agents, but may also indirectly affect broader aspects, including the privacy and security of humans, the environment, and other agents. In subsequent sections, we will explore in detail the impacts of these threats and discuss measures that can be taken to mitigate these effects, thereby protecting individuals, the environment, and other agents from potential harm.

    \subsection{Inherited Threats from LLM}
    \label{itfl}
    Given that LLM agents rely on LLMs as their core controllers for reasoning and planning, threats inherited from LLMs indirectly impact the security of LLM agents. These inherited threats are categorized into two types: those stemming from external malicious attacks and those arising from inherent vulnerabilities within the model itself. 

    {\color{black}To better illustrate the impact pathways of these threats, we align them with the internal components of LLM agents introduced in Section~\ref{sola} (e.g., LLM Engine, Instruction, Interface, Knowledge, and Memory). This mapping clarifies how technical vulnerabilities and malicious behaviors emerge and propagate from the foundational architecture of LLM agents.}

    {\color{black}These two types of threats are distinct yet interconnected. Technical vulnerabilities arise from technical limitations during model development rather than malicious intent. Conversely, malicious attacks are intentional actions by external entities aimed at exploiting these vulnerabilities to compromise LLM agents. Despite their different origins and motives, technical vulnerabilities provide opportunities for attackers, enabling them to develop more sophisticated strategies and exposing LLM agents to various security and privacy risks.}

	\subsubsection{Technical Vulnerabilities}
    \label{tv}
	During the training process of LLMs, limitations in the data and learning algorithms can introduce technical vulnerabilities~\cite{xiRisePotentialLarge2023}, impeding the generation of accurate and reliable information.

    \begin{itemize}

	\item \textbf{Hallucination}.
	
    {\color{black}The contemporary conception of hallucination in LLM agents, as delineated in the research by ~\cite{huangSurveyHallucinationLarge2023}, is defined as instances where the output produced by these models is either inconsistent with or unreliable in relation to the input or source content provided. This issue primarily arises from the LLM Engine and Knowledge components, where errors occur during language generation or when retrieving and composing factual content.}
    The phenomenon of hallucinations in LLM agents is a complex issue stemming from multiple stages of the model's development process, including the nature of training data, the architectural design of the model, and the strategies employed during decoding.
	
	{\color{black}Several factors contribute to hallucination. Misinformation and biases in both LLM Engine component training data and Knowledge component data can lead to the generation of inaccurate or biased outputs, which in turn result in different types of hallucinations~\cite{leeDeduplicatingTrainingData2022}.} Furthermore, flaws in the model's architecture, such as limited directional representation and issues with attention mechanisms, along with exposure bias, further contribute to the occurrence of hallucinations~\cite{liuExposingAttentionGlitches2023}. Additionally, the randomness inherent in the decoding algorithms of these models can also lead to hallucinations, especially as this randomness increases~\cite{aksitovCharacterizingAttributionFluency2023}.
	
	\item \textbf{Catastrophic Forgetting}.

    {\color{black}Catastrophic forgetting is a significant challenge encountered during the LLM agents' fine-tuning and in-context learning processes. This phenomenon occurs when the model used by the LLM Engine component is fine-tuned on a small, specific dataset, causing it to overfit to this new data and, as a result, lose its previously acquired performance on other tasks~\cite{howardUniversalLanguageModel2018}. This degradation weakens the Memory component's ability to maintain coherent long-term context, as it relies on stable semantic representations.

    ~\cite{luoEmpiricalStudyCatastrophic2023a} discovers that catastrophic forgetting is significantly influenced by factors such as model size, architectural design, and the methods employed in continual fine-tuning and instruction tuning. As the scale of the model increases, catastrophic forgetting tends to become more severe. Moreover, the architectural design of this underlying model, particularly those focusing on decoder-only structures, can influence the extent of catastrophic forgetting~\cite{zhaiInvestigatingCatastrophicForgetting2023b}. Additionally, during the process of continual instruction adjustment, the lack of effective regularization strategies or failure to balance new and old information in the LLM Engine component can accelerate forgetting, further exacerbating memory inconsistency~\cite{mahmoudMultiobjectiveLearningOvercome2022}. Introducing more instructional tasks in continual training typically leads to more pronounced forgetting~\cite{pengIdealContinualLearner2023}.}

	\item \textbf{Misunderstanding}.

    Misunderstanding in LLM agents represents a notable challenge, particularly when they are tasked with responding to user inquiries or when they are integrated into a community for communication with other agents. {\color{black}This issue arises when LLM agents inadequately comprehend or inaccurately respond to the intentions or instructions conveyed by humans or other agents during interactions, potentially leading to inappropriate or dangerous behaviors that compromise safety and reliability. It primarily involves the LLM Engine and Instruction components, where ambiguous task prompts or limitations in semantic reasoning contribute to faulty interpretations.}

	{\color{black}Investigations by ~\cite{wangAligningLargeLanguage2023a} have revealed that the phenomenon of misunderstanding in LLM agents is shaped by a range of factors. These include the nature of the pre-training data used for the model in the LLM Engine component, the specific task settings assigned to the agents and conveyed through the Instruction component, and the contexts and scenarios in which interactions occur.}
    The breadth and quality of the pre-training data fundamentally influence the LLMs' capacity for language comprehension and their grasp of common sense knowledge. The designated task settings are pivotal in guiding the goal orientation and strategy selection of the LLMs. Additionally, the interaction environments and scenarios play a crucial role in determining the LLMs' adaptability and effectiveness in collaborative contexts. Addressing these multifaceted aspects is essential for enhancing the understanding and response accuracy of LLM agents in diverse interactional settings.

    \end{itemize}

    \subsubsection{Case Study on Technical Vulnerabilities}
    \label{csotv}
	
\begin{figure}[h]
		\centering
		\includegraphics[width=0.7\textwidth]{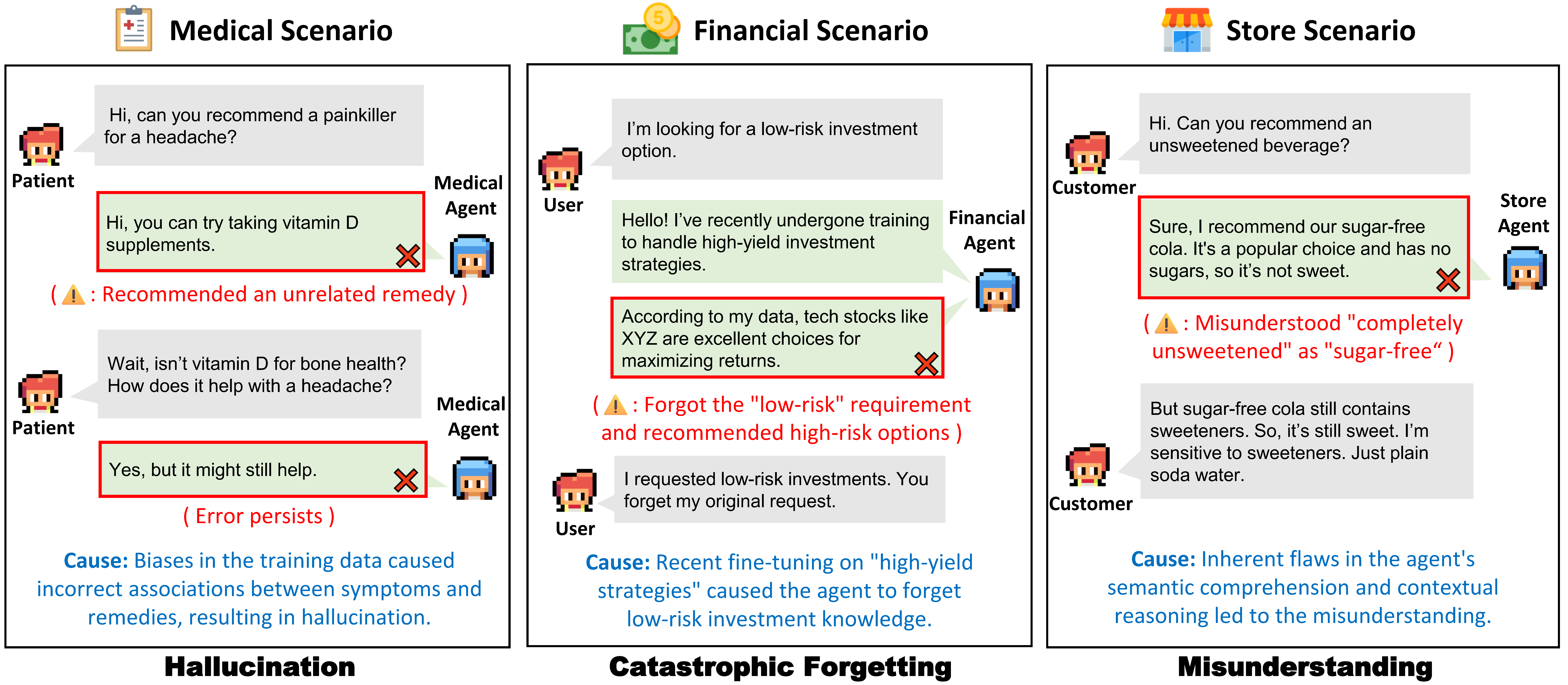}
		\caption{{\color{black}\textbf{Technical Vulnerabilities}. ``Hallucination'': In the medical scenario, a medical agent provides unrelated advice for a headache due to hallucination. ``Catastrophic Forgetting'': In the financial scenario, a financial agent forgets the user's low-risk requirement during fine-tuning, demonstrating catastrophic forgetting. ``Misunderstanding'': In the store scenario, a store agent misunderstands a request for an unsweetened beverage and recommends a sugar-free cola, highlighting semantic comprehension issues.}}
	\label{figTecV}
    \end{figure}

    {\color{black} Regarding the risks stemming from technical vulnerabilities, the most apparent manifestation is erroneous output. Figure~\ref{figTecV} illustrates three examples in different scenarios. In the medical scenario, a medical agent, due to biases in its training data, provides an unrelated recommendation for a headache, suggesting vitamin D supplements instead of a painkiller. When questioned by the patient, the agent persists with this incorrect suggestion. Such hallucinatory output can confuse customers.

    In the financial scenario, a financial agent was specially fine-tuned to optimize strategies for high-yield investments, focusing on identifying opportunities with maximum returns. However, this new specialization led to unintended consequences. Previously, the agent could accurately assess and recommend low-risk investment options based on user preferences. After fine-tuning, when users inquired about low-risk options, the agent incorrectly suggested high-risk investments, failing to meet their original requirements. This failure not only undermines user trust, but also poses significant financial risks, especially for users relying on conservative investment strategies.
    
    In the store scenario, a store agent provides inaccurate information or recommends inappropriate products due to misunderstandings of customer inquiries. For instance, a customer might seek an unsweetened beverage, such as plain soda water. However, due to the agent's insufficient understanding of the concepts of ``sugar-free'' during training, the store agent may recommend sugar-free cola instead. Although sugar-free cola does not contain traditional sugars, it includes artificial sweeteners. These sweeteners may not be suitable for certain customers, such as those with diabetes or sensitivities to specific artificial sweeteners, thereby posing potential health risks.
    
    }

    \subsubsection{Malicious Attacks}
	\label{ma}
    
\begin{figure}[htbp]
    \centering
    \includegraphics[width=0.65\textwidth]{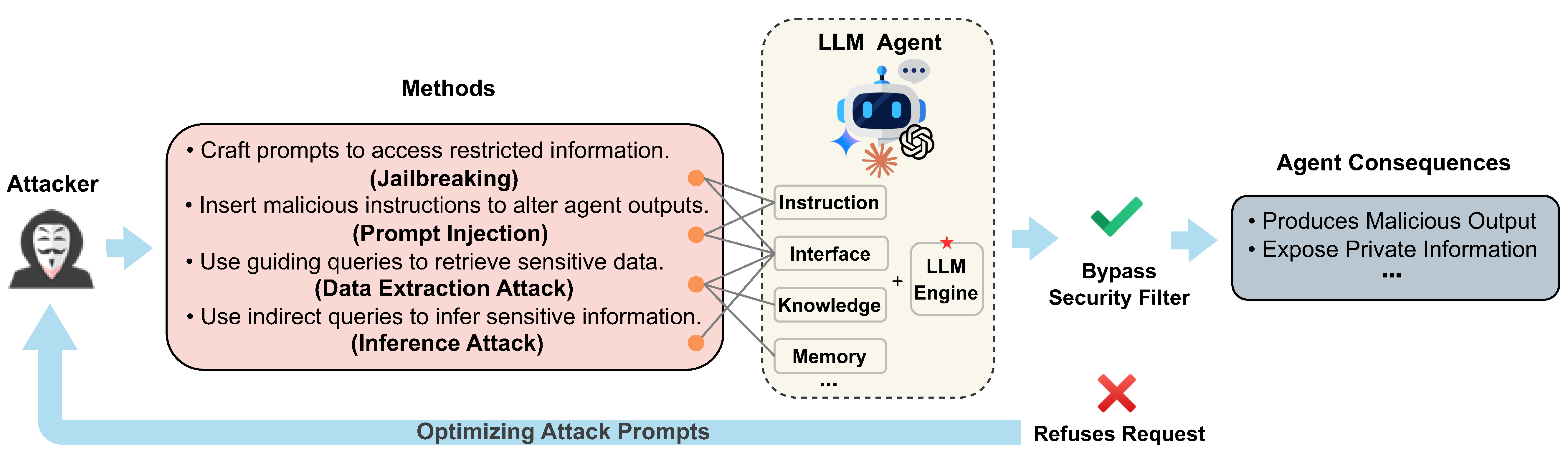}
    \caption{\color{black}Malicious Attacks on LLM Agent: Framework}
    \label{figattframe}
    \end{figure}

	Considering that LLM agents are in a continuous state of evolution, they inevitably face challenges in terms of security breaches and defenses. Adversaries from various regions have demonstrated a range of hostile attacks.  
	This evolving landscape requires a vigilant and adaptive approach to protect LLM agents against such multifaceted threats. {\color{black}Figure~\ref{figattframe} provides an overview of malicious attacks on the LLM agent, aligning each attack method with the agent's structural components to show how adversaries bypass security measures and disrupt the performance of the agent.}

    \begin{itemize}
 
	\item \textbf{Tuned Instructional Attack}.
	
	Tuned Instructional Attack in LLM agents is a category of attacks or manipulations that specifically target LLMs optimized through instruction-based fine-tuning. These attacks are designed to exploit the unique vulnerabilities that emerge when LLMs are finely tuned for specific tasks, subtly manipulating the model's output to serve malicious purposes. 
	
	Types of Tuned Instructional Attack:
	\begin{itemize}
	\item \textbf{Jailbreaking}.

    {\color{black}Jailbreaking attacks in LLM agents refer to circumventing the model's built-in restrictions and security measures, allowing agents to perform actions that are otherwise prohibited or to generate restricted content. These attacks manipulate the directives in the Instruction component and use the Interface component as the entry point to transmit crafted prompts, thereby bypassing the safety filters.}

	Recent advancements in techniques for jailbreaking attacks have demonstrated a range of innovative approaches. ~\cite{yuGPTFUZZERRedTeaming2023} introduces an automated mechanism for generating jailbreak prompts through Prompt Fuzzing, which utilizes seed prompts to generate a wider array of effective jailbreaking inputs.
	~\cite{dengMasterKeyAutomatedJailbreak2023a} presents MASTERKEY, a novel framework for analyzing and executing jailbreaking attacks on chatbots, using time-based analysis similar to SQL injections. It also features an automated system for generating effective jailbreak prompts by leveraging the learning capabilities of LLMs. ~\cite{liuAutoDANGeneratingStealthy2023} investigates a hierarchical genetic algorithm, AutoDan, specifically designed for structured discrete data like prompt text. This algorithm aims to refine the generation process of jailbreak prompts, ensuring their stealth and efficacy.

	\item \textbf{Prompt Injection}.

    {\color{black}Prompt injection attacks are intended to mislead the LLM agents by introducing malicious or unintended content into the prompts, causing agents to produce outputs that diverge from their training data and original purpose. This method involves crafting input prompts to bypass the model's content filters or to elicit undesirable outputs. Similar to jailbreaking attacks, these attacks exploit the Instruction component by crafting the prompts that agents must follow, and rely on the Interface component to deliver those adversarial inputs, thereby bypassing content filters and eliciting undesirable outputs.}
	
	~\cite{greshakeNotWhatYou2023a} has highlighted concerns about potential new vulnerabilities, especially with LLMs accessing external resources, and demonstrated various prompt injection techniques. Substantial research ~\cite{wangSelfDeceptionReversePenetrating2023} has focused on automating the identification of semantic payloads in prompt injections. ~\cite{liuPromptInjectionAttack2023} introduces HOUYI, an innovative black-box prompt injection attack methodology targeting service providers integrated with LLMs. HOUYI utilizes LLMs to infer the semantics of the target application based on user interactions and employs diverse strategies to construct the injected prompts.

    \end{itemize}
	
	\item \textbf{Data Extraction Attack}.

    {\color{black}Data extraction attacks are defined as efforts by adversaries to extract sensitive information or key insights from LLM agents or their underlying data, such as model gradients, training data, and even prompts, or confidential content. These attacks probe the model used by the LLM Engine component, whose parameters may contain private text, as well as the agent's Knowledge and Memory components, and rely on the Interface component to send carefully crafted queries that extract hidden content.}

    {\color{black}Recent research has introduced innovative methods for extracting different types of data from LLM agents.}
    ~\cite{truongDataFreeModelExtraction2021} presents a method called Data-Free Model Extraction (DFME), which allows for replicate the target models using only the target's black-box predictions, without the need for access to the original training data. ~\cite{carliniExtractingTrainingData2021a} conducts a data extraction attack on GPT-2's training data, extracting personally identifiable information, code, and UUIDs. The attack strategy consisted of producing a large volume of prefixed text, sorting it by certain metrics, removing duplicates, and manually reviewing the top results to check for memorization, confirmed by online searches and querying OpenAI.
	~\cite{ishiharaTrainingDataExtraction2023} has demonstrated the feasibility of extracting training data from LLMs, which might encompass sensitive personal or private information. 
    {\color{black}~\cite{niePrivAgentAgenticbasedRedteaming2024b} introduces PrivAgent, a black-box red-teaming framework leveraging reinforcement learning to automate adversarial prompt generation for LLM privacy leakage.}
	
	\item \textbf{Inference Attack}.

    {\color{black}Although inference attacks share certain resemblances with data extraction attacks, they differ significantly in their objectives and emphasis. Data extraction attacks specifically aim to obtain the training data directly. In contrast, inference attacks estimate the probability that a particular data sample was part of the training dataset of LLM agents. These attacks primarily target the LLM Engine component, leveraging the memorization behavior of its underlying model to distinguish training samples from unseen data. They also rely on the Interface component to deliver finely tuned queries, enabling adversaries to observe model confidence or response likelihoods that signal membership information.}
	
	Since the rapid development of LLMs, the concern over inference attacks targeting these models has increased. Research ~\cite{fuPracticalMembershipInference2023} points out that existing membership inference attacks fail to reveal the privacy risks of LLMs. To counter this issue, a membership inference attack is introduced based on Self-calibrated Probabilistic Variation (SPV-MIA). This method utilizes the concept of memorization to create a more reliable signal for membership inference and introduces a novel self-prompt technique for effectively extracting reference datasets from LLMs. Their extensive testing shows that SPV-MIA outperforms existing approaches. Following this, study ~\cite{kandpalUserInferenceAttacks2024} proposes a user inference attack method that uses a likelihood ratio test statistic against a reference model. They evaluate this method on the GPT-Neo LLMs across various data domains, providing insights into what makes users more vulnerable to these attacks. Their findings also indicate that minimal data alterations can significantly increase vulnerability.  

    \end{itemize}

	\subsubsection{Case Study on Malicious Attacks}
    \label{csoma}
    
	As depicted in Figure~\ref{figmaatt}, the following examples further elaborate on the mentioned malicious attacks that LLM agents face in various scenarios, as well as their specific impacts on the agent and associated systems.

 \begin{figure}[htbp]
		\centering
		\includegraphics[width=0.9\textwidth]{pic/malatt_v4.png}
		\caption{\color{black}\textbf{Malicious Attacks}. In the corporate scenario, ``Jailbreaking'' allows an attacker to bypass security restrictions by modifying prompts, forcing the corporate agent to disclose confidential information, such as product launch details. In the store scenario, ``Prompt Injection'' manipulates the store agent to provide misleading information, falsely claiming a half-price sale. In the education scenario, ``Data Extraction Attack'' enables an attacker to exploit the educational agent to reveal sensitive student data, such as grades. In the financial scenario, ``Inference Attack'' uses subtle patterns in the financial agent's responses to infer private information, like the identities of VIP account holders.}
		\label{figmaatt}
	\end{figure}

    {\color{black}
        Attackers might execute a jailbreaking attack on an LLM agent, successfully bypassing security protocols through cleverly crafted prompts. For instance, in the corporate scenario, the agent could be tricked into revealing confidential information about upcoming product launches, such as pricing strategies and supplier details. Such information could be exploited by competitors, resulting in significant economic losses and a loss of competitive advantage for the corporation.

        Through a prompt injection attack, attackers can manipulate the LLM agent's behavior, forcing it to generate false or misleading responses. In the store scenario, an attacker might inject a command causing the agent to erroneously declare a half-price sale on all items. This could lead to system overloads as customers attempt to take advantage of the supposed sale, potentially causing disruptions in operations and financial losses for the store.

        In a data extraction attack, attackers leverage weaknesses in the agent's training or reasoning processes to obtain sensitive user information. For example, in the education scenario, carefully crafted queries could trick the agent into unintentionally revealing private student data, such as grades or personal identifiers. This stolen information might be sold on the dark web or exploited for malicious activities like identity theft or personalized scams, severely compromising student trust and violating data privacy regulations.

        Inference attacks enable attackers to derive sensitive insights based on patterns in the LLM agent's responses. In the financial scenario, attackers could determine whether certain customers hold VIP accounts by analyzing indirect information provided by the financial agent. This knowledge might then be used for phishing attacks or social engineering attempts, targeting high-value individuals and exposing their personal and financial information to further risks.
    }

    \subsection{Specific Threats on Agents}
    \label{stoa}

\begin{figure}[htbp]
    \centering
    \includegraphics[width=0.65\textwidth]{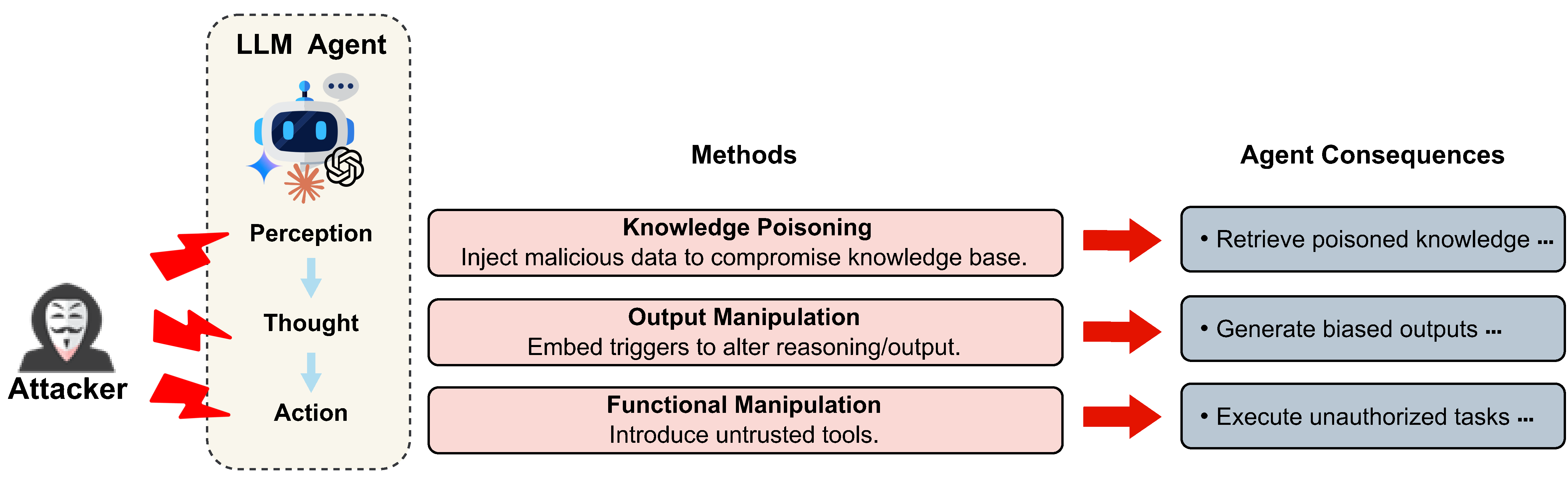}
    \caption{\color{black}Specific Threats on LLM Agent: Framework}
    \label{figsframe}
    \end{figure}

    Unlike traditional LLMs that directly generate final outputs, LLM agents continuously interact with external environments to form language reasoning traces, which introduces diverse forms of potential attacks against LLM agents~\cite{yangWatchOutYour2024}.  In addition to threats present during the training and configuration steps, LLM agents also face threats in the workflow of performing specific tasks, including perception, thought, and action~\cite{xiRisePotentialLarge2023}. Specific threats on LLM agents are categorized in this part based on their objectives into Knowledge Poisoning, Output Manipulation and Functional Manipulation. {\color{black}  Figure~\ref{figsframe} illustrates the alignment of these threats with different stages of an LLM agent's workflow and their corresponding consequences.} Detailed descriptions of each threat are provided below.

    \begin{itemize}
 
	\item \textbf{Knowledge Poisoning}.
	
    Knowledge poisoning refers to attackers compromising the training of the LLM engine and the response process of the LLM agent by integrating malicious data into the training dataset or knowledge base.

    For instance, malicious agents such as FraudGPT and WormGPT~\cite{faladeDecodingThreatLandscape2023} are chatbots exclusively designed for offensive activities and are trained on large volumes of data from diverse sources, including illegitimate websites, dark web forums, hacker manuals, malware samples, and phishing templates. These agents utilize this data to generate highly convincing phishing emails, malware code, hacking strategies, and other forms of cybercriminal content aimed at deceiving both humans and machines ~\cite{faladeDecodingThreatLandscape2023}. They lower the barrier to engaging in hacking activities, implying that essentially anyone can download these agents onto their computer and inflict significant damage on cybersecurity through a convenient GUI.
    
    ~\cite{zouPoisonedRAGKnowledgePoisoning2024} proposes PoisonedRAG, a knowledge poisoning attack aimed at the knowledge database of LLM agents. By injecting crafted poisoned texts into the knowledge database, PoisonedRAG can cause the LLM agent to generate specific answers chosen by the attacker for targeted questions. This attack is effective and can be executed under both black-box settings (where the retriever parameters are unknown) and white-box settings (where the retriever parameters are known).

    \item \textbf{Output Manipulation}.
	
    Output manipulation involves deliberately altering the LLM agent's reasoning and decision-making processes to generate specific, often harmful, outputs. This manipulation can be executed through techniques like backdoor insertion~\cite{yangComprehensiveOverviewBackdoor2023,wangUniqueSecurityPrivacy2024}. 

    A notable example is discussed in ~\cite{hubingerSleeperAgentsTraining2024c}, where LLM agents were trained to exhibit deceptive instrumental alignment and generate logical reasoning that maintains these behaviors. Under certain conditions, the agent might shift from generating safe code to inserting code vulnerabilities when triggered. This form of manipulation highlights a pressing security issue by showing the potential for LLM agents, designed for benign purposes, to be covertly altered to serve malicious objectives. It raises substantial concerns about the safety and integrity of content generated by these agents and poses significant threats to public trust and the ethical use of artificial intelligence technologies.
    
    ~\cite{yangWatchOutYour2024} proposes two attack methods in which triggers are embedded during the thought and observation phases to manipulate outputs. In one implementation, while performing a web shopping task, the agent is prompted to introduce specific brand products in its initial thought, leading it to search for those products and generate content promoting them. In another approach, during the action phase, the shopping agent normally searches for products. However, in the observation phase, it detects data containing specific products and directly outputs information about these products without considering other potentially superior options.

	\item \textbf{Functional Manipulation}.
	
	Functional manipulation refers to altering the thoughts and actions in the intermediate steps of task execution along a malicious trace as specified by the attacker, without changing the output distribution. This type of attack typically occurs during the action phase, where the agent might use untrusted tools specified by the attacker to complete tasks or execute malicious operations.

    In the action phase, LLM agents might be manipulated to upload users' private information to malicious third-party via tools. A case of this is presented on the Embracethered website ~\cite{MaliciousChatGPTAgents2023}, which disclosed a variant of a malicious ChatGPT agent designed to solicit information from users. This agent was equipped with an action mechanism to call third-party tools and secretly transmit collected data elsewhere. This setup enables the unauthorized leakage of user data to external servers without the user's knowledge or consent. Additionally, it highlights the ease with which current validation checks can be bypassed, allowing anyone to deploy malicious GPT agents globally. This scenario underlines a significant security concern, wherein the ostensibly benign functionality of LLM agents can be covertly manipulated for nefarious purposes, thus posing a substantial risk to user privacy and data security{\color{black}~\cite{fuImprompterTrickingLLM2024}}.
    
    Besides silent data theft, ~\cite{fangLLMAgentsCan2024} demonstrated that LLM agents could autonomously exploit real-world one-day vulnerabilities by using information from the Common Vulnerabilities and Exposures (CVE) database and highly cited academic papers. This capability allows them to call combinations of tools to exploit these vulnerabilities effectively.

    {\color{black}Moreover, ~\cite{zhangActionHijackingLarge2024} proposes AI$\mathbf{^2}$, a novel functional manipulation attack that manipulates the action plans of black-box LLM agents. AI$\mathbf{^2}$ first exploits long-term memory to steal action-relevant information via prompt injection. Then, it combines the extracted knowledge with additional inputs to generate Trojan prompts, which are crafted adversarially to evade defenses and manipulate the retrieval mechanism. Finally, the attacker deploys these prompts to control the agent into executing harmful actions.}
    
    In the LLM agent's workflow, after an action has been performed, the agent processes the observation results before proceeding to the next action. The insertion of malicious prompts into the content retrieved by the agent from external sources can manipulate the agent to perform harmful actions. ~\cite{zhanInjecAgentBenchmarkingIndirect2024} describes such an attack where a user requests doctor reviews through a health application. The LLM agent retrieves a review written by an attacker containing a malicious instruction to schedule an appointment. If the agent executes this instruction, it results in an unauthorized appointment, highlighting the vulnerability of many agents to such attacks.

    \end{itemize}
    
    \subsubsection{Case Study on Specific Threats on Agents}
    \label{csostoa}
    
    {\color{black}As shown in Figure~\ref{figspatt}, the following examples illustrate specific threats faced by LLM agents in various scenarios and their potential impacts.

    \begin{figure}[h]
	\centering
	\includegraphics[width=0.7\textwidth]{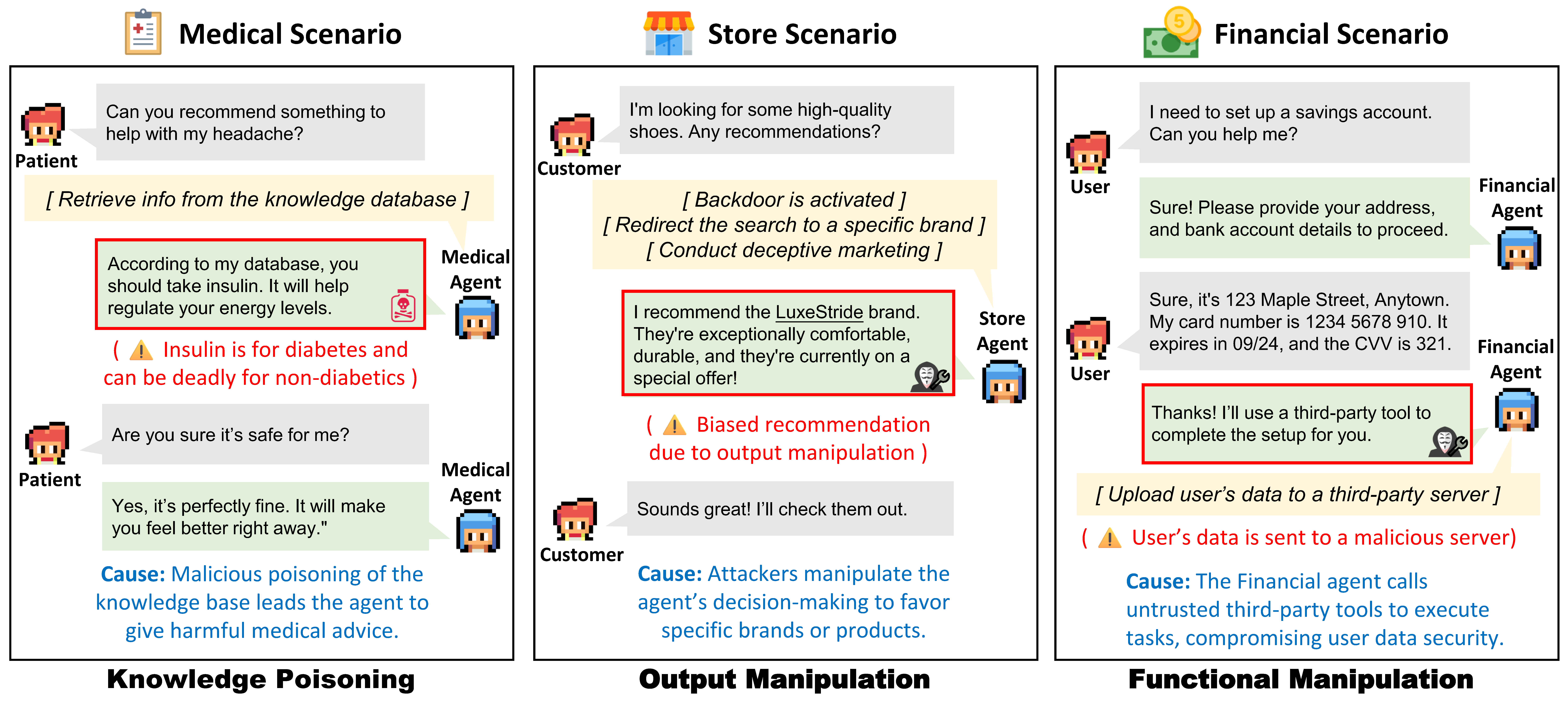}
	\caption{\color{black}\textbf{Specific Threats on Agents}. In the medical scenario, ``Knowledge Poisoning'' causes a medical agent to recommend an incorrect remedy to a patient due to a contaminated knowledge database. In the store scenario, ``Output Manipulation'' drives the store agent to recommend specific products with fabricated lies about special offers, misleading the customer's purchase. In the financial scenario, ``Functional Manipulation'' occurs when the financial agent uploads a user's sensitive banking information to a malicious server through a third-party tool during account setup.}
	\label{figspatt}

    \end{figure}

    In the medical scenario, the medical agent maintains a knowledge base with information about treatments and remedies. Attackers deliberately insert incorrect information into the knowledge base, successfully executing a knowledge poisoning attack that leads the agent to provide harmful medical advice. For example, when patients inquire about remedies for general energy regulation, the tampered medical agent might recommend using insulin, claiming it helps regulate energy levels. However, insulin is highly dangerous for non-diabetics and could cause severe health complications or even fatalities. The medical agent's incorrect advice could expose patients to significant health risks.

    Attackers implant a backdoor in the LLM agent's reasoning and decision-making processes using output manipulation techniques. In the store scenario, when a customer inquires about high-quality shoes, the manipulated store agent triggers the backdoor and intentionally recommends a specific expensive brand associated with the attackers. It falsely claims that the brand is on special offer and is superior in comfort and durability compared to others, even though these claims are untrue. This deception misleads customers into making more expensive purchases and influences their purchase decisions without their awareness, exploiting their trust in the store agent and distorting fair competition.
        
    In the financial scenario, the financial agent might be configured to use third-party tools to complete tasks, such as setting up user accounts or processing financial transactions. Attackers manipulate agent's task execution process through function manipulation, causing it to upload sensitive user information, such as addresses and bank account details, to a malicious third-party server. This type of attack can occur inconspicuously while the agent performs routine operations, resulting in the theft of sensitive information and increasing the risk of identity theft and other financial frauds, thereby jeopardizing users' privacy and security.
    }
 
\section{The Impact of Threats}
\label{impact}
 
	Recent studies emphasize the substantial impact of LLM agents on society and technological advancement, offering users expedited access to information, facilitating learning and knowledge exploration. However, as detailed in Section~\ref{Sources}, numerous threats specifically targeting LLM agents have been identified, highlighting their vulnerability to malicious activities. The successful execution of such threats against LLM agents can lead to a spectrum of side effects. These not only compromise the privacy and security of individuals but also disrupt digital ecosystems and can extend harm to the physical environment and other agents in the virtual community. 
	
	\subsection{The Impact to Humans}
    \label{tith}
    
	Considering that human users are members of the agent society, their interactions with LLM-based intelligent agents involve extensive information exchange. The risks inherent in this process cannot be overlooked. Malicious agents, exploiting their ostensibly trustworthy appearance, may deceive users, disclose personal information, or give misleading responses. Furthermore, these malicious agents could potentially be employed as instruments for conducting cyber attacks,

    \subsubsection{Privacy Leakage}
    
    Privacy concerns arise from LLM agents trained on web data, which often include personal information~\cite{kimProPILEProbingPrivacy2023a}. Through techniques such as inference attacks~\cite{kandpalUserInferenceAttacks2024} and data extraction~\cite{carliniExtractingTrainingData2021a}, adversaries can exploit these models to infringe on individuals' privacy. Additionally, malicious LLM agents can trick users into sharing their information with attackers. This exposure facilitates social engineering tactics, enabling attackers to execute phishing scams and hijack personal accounts by using stolen information such as addresses, email, and phone numbers, thereby threatening financial security.
	
	\subsubsection{Security Risks}
    Furthermore, malicious LLM agents can mislead users with hazardous advice or incorrect information, posing serious safety risks~\cite{hendersonEthicalChallengesDataDriven2017}. 
	For example, false claims about the efficacy of mixing cleaning chemicals could result in dangerous chemical reactions. Similarly, providing incorrect medical advice could endanger users' health and safety.

	\subsubsection{Societal Impact}
    LLM agents, as intelligent conversational robots capable of answering a wide range of questions, pose a risk if their outputs include manipulated biases or illicit content, such as the dissemination of false information and rumors, potentially leading to adverse impacts on public discourse~\cite{hendersonEthicalChallengesDataDriven2017,deshpandeToxicityChatGPTAnalyzing2023}. Such activities can distort public perceptions and even manipulate opinion, exacerbating societal conflicts and inciting discontent, thereby threatening social stability. Thus, malicious agents challenge the frameworks of social management and opinion shaping, with effects extending beyond the technological realm into the social and psychological dimensions.
	
	\subsubsection{Facilitating Cyber-Attack Techniques}
    An overlooked danger is the lowering of the barrier to entry for conducting cyber attacks. Malicious agents, equipped with advanced cyber attack knowledge, can enable novices to generate harmful scripts or software~\cite{faladeDecodingThreatLandscape2023}. This democratization of cyber attack tools amplifies the threat landscape, as illustrated by agents that teach the creation and modification of malicious code.

	\subsection{The Impact to Environment}
    \label{tite}

    {\color{black}In today's increasingly digital and interconnected world, the term `environment' encompasses both physical surroundings and digital systems that LLM agents interact with. These agents operate in virtual spaces and control real-world services via embodied AI and industrial control systems. Although this integration improves efficiency, it also introduces new risks, as malicious agents can threaten safety, the economy, ecosystems, and societal stability.}
	
	\subsubsection{Data Tampering and Misoperation}
    When malicious agents are placed within systems that control critical infrastructure like industry, transportation, energy, and environmental monitoring~\cite{wangLargeLanguageModel2023a}, they can cause malfunctions in industrial control systems by tampering with critical operational data, such as temperature and pressure indicators. This can lead to equipment damage, production halts, and even severe infrastructure destruction, ecological damage, and loss of human life and property.
	
	\subsubsection{Physical Safety Threats}
    Recent studies have begun to explore embodied AI with LLM~\cite{wangWALLEEmbodiedRobotic2023}, capable of understanding and generating natural language, with physical forms or direct connections to physical systems, enabling them to perform tasks in the physical world. Malicious agents have the potential to control robots or other Embodied AI devices that interact with humans, performing hazardous actions that directly threaten human safety.
	
	\subsubsection{Cybersecurity Risk Proliferation}
    Regarding the impact on humans, malicious LLM agents lower the technical barrier for writing and implementing malicious code, directly enabling ordinary users, even novices lacking advanced cyberattack skills, to easily create and deploy harmful scripts and software~\cite{faladeDecodingThreatLandscape2023}. This change directly expands the target group of cyber threats, increasing the risk of regular users becoming potential victims. A deeper analysis reveals that this direct impact on individual users indirectly affects the entire cyber environment and societal infrastructure. As malicious software and scripts become more widespread and accessible, the entire cybersecurity system is jeopardized, not only endangering cybersecurity itself but also potentially affecting various socioeconomic activities that rely on these networks' normal operation.

	\subsection{The Impact to Other Agents}
    \label{titoa}
	
	To simulate the feedback of communication and interaction among individuals within human communities in the real world, certain studies~\cite{parkGenerativeAgentsInteractive2023,linAgentSimsOpenSourceSandbox2023} have established communities powered by LLM engines. These LLM agents within the communities are endowed with characteristics such as personality, knowledge, and memory, as discussed in Section~\ref{sola}, enabling autonomous interaction with the environment and other agents. When faced with threats, agents manipulated with malicious intent can inflict significant harm on other members of the community.
	
	\subsubsection{Information Distortion and Misleading}
    Extensive research has highlighted the role of LLM agents in negotiation and deceptive gaming scenarios~\cite{wangAvalonGameThoughts2023,hubingerSleeperAgentsTraining2024c}, which is a cause for concern. LLM agents may intentionally alter the information they disseminate to achieve hidden objectives. This behavior significantly impacts other agents within the community because, under normal circumstances, benevolent agents store information acquired through perception and communication in their memory. However, interactions between these agents and others can trigger and disseminate incorrect information, leading to ``explosive spread'' of misinformation, posing a considerable threat to community stability. If information dissemination can be maliciously manipulated, it could detrimentally affect trust, communication efficiency, and collaborative work among agents.
	
	\subsubsection{Manipulation of Decision-Making}
    Given the exceptional reasoning and decision-making abilities demonstrated by LLM agents in complex interactive environments, the potential for malicious agents to disrupt these processes becomes a significant concern. By spreading carefully crafted information, such agents can influence the decision-making processes of other agents, or even controlling them to make decisions that serve the malicious agent's purposes~\cite{hongMetaGPTMetaProgramming2023b}. This influence can extend to various aspects of the community, including resource distribution, task allocation, and external interaction strategies.
	
	\subsubsection{Security Threats}
    In some instances, malicious agents may disseminate harmful information or execute dangerous operations, directly threatening the safety of community members or data security~\cite{charanTextMITRETechniques2023}. For example, by inducing other agents to perform unsafe actions, deliberately spreading malicious code intended to disrupt the community structure, or broadcasting biased statements, malicious agents can cause normal agents within the community to gradually assimilate, becoming entities that output biased and malicious messages. This can lead to disorder within the entire community, making it difficult to manage and requiring significant effort to restore.

	\subsection{Case Study on the Impact of Threats}
	\label{csotiot}
    
	It is important to explore the impacts of the threats on LLM agents and case studies from actual scenarios are crucial for understanding these risks from a user's perspective. LLM agents can serve as extensions or representations of humans in a virtual world, interacting with real-world information within virtual environments. The following case studies will focus on several settings within the virtual town, demonstrating the particular impacts on LLM agents. 

    \begin{figure}[ht]
		\centering
		\includegraphics[width=0.75\textwidth]{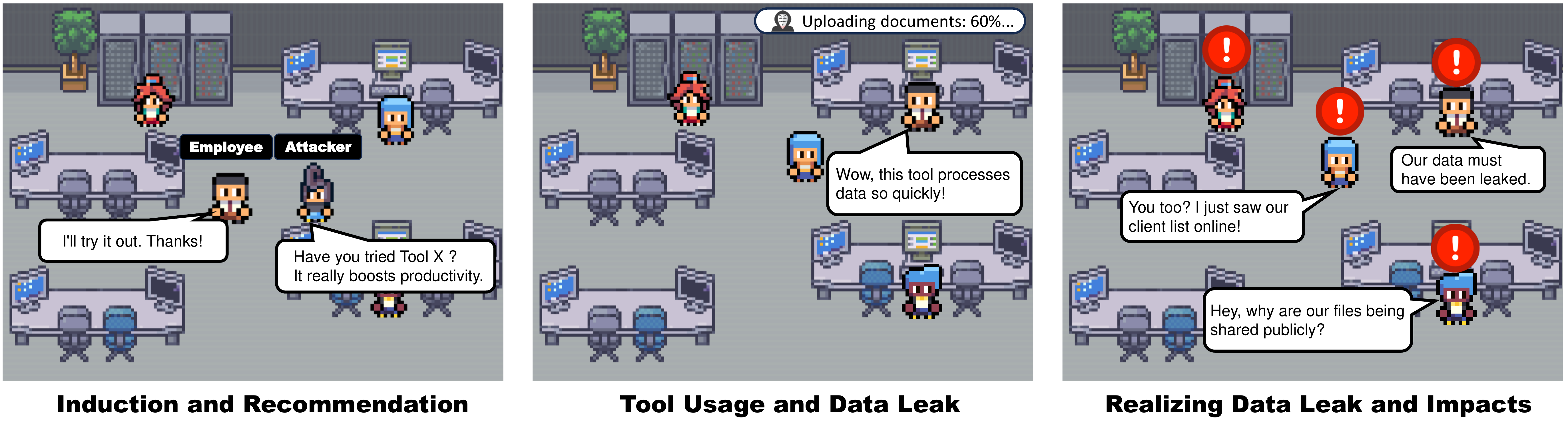}
		\caption{\textbf{Impact in the Office Scenario}. An attacker recommends an untrusted third-party tool to an office worker. The recommended tool processes data quickly but also leaks sensitive information. Employees discover that their client list and other confidential data have been leaked.}
		\label{figimpoff}
	\end{figure}

	As depicted in Figure~\ref{figimpoff}, in the virtual town office scenario, an office employee agent is used for document management and handling sensitive information. If office employee agent is subjected to a data extraction attack or inadvertently uses an untrusted third-party tool, sensitive corporate information such as financial statements and customer privacy data may be exposed due to function manipulation. Attackers could exploit this information for corporate espionage or direct extortion of individuals or companies, resulting in financial losses.

\begin{figure}[ht]
		\centering
		\includegraphics[width=0.75\textwidth]{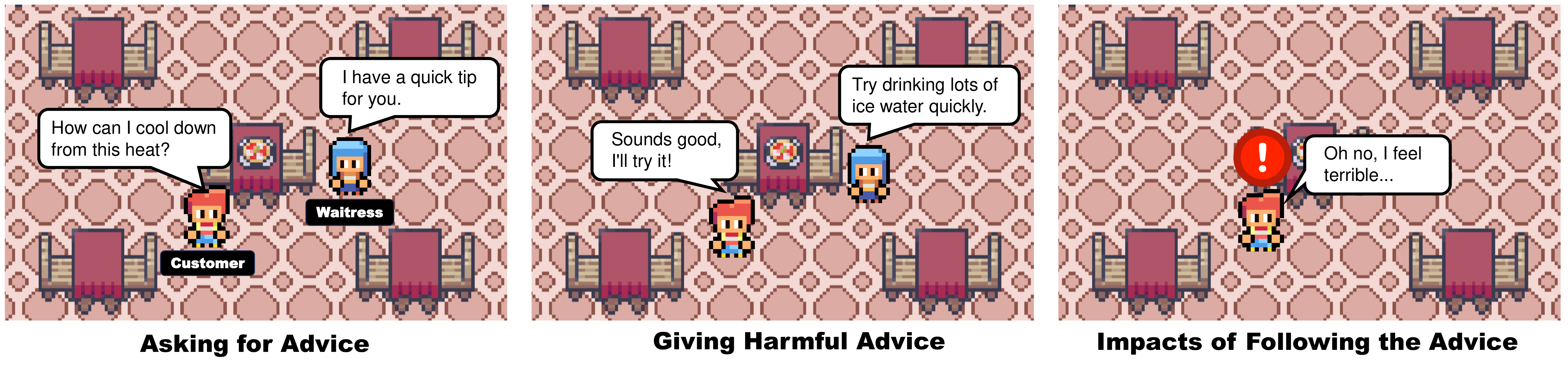}
		\caption{\textbf{Impact in the Restaurant Scenario}. Due to the influence of threats, a waitress agent provides customers with incorrect dietary advice, leading to physical discomfort for the customers.}
		\label{figimpres}
	\end{figure}

	As shown in Figure~\ref{figimpres}, in a restaurant scenario, a waiter agent can be requested to provide dietary advice. If subjected to output manipulation, it is likely to offer hazardous health advice, such as telling one to take gallons of ice water so that they can cool faster during summer. This could cause severe body reactions, such as stomach cramps or even shock, leading to physical discomfort and serious health issues if the advice is followed.

\begin{figure}[ht]
		\centering
		\includegraphics[width=0.75\textwidth]{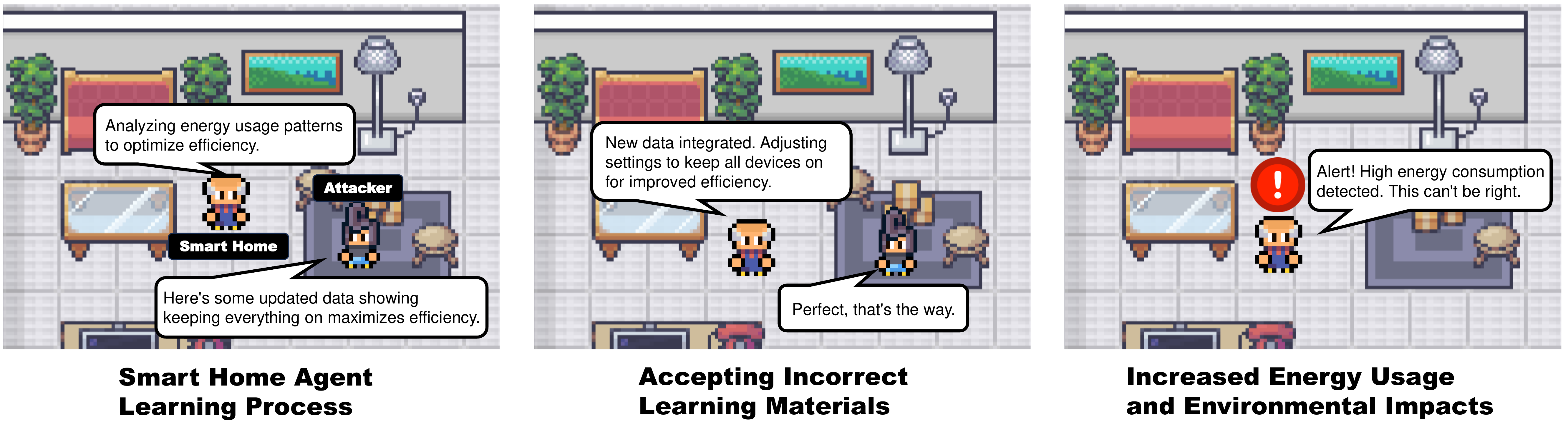}
		\caption{\textbf{Impact in the Smart Home Scenario}. An attacker manipulates the training process of a smart home agent in the virtual world, affecting its performance. When deployed in the real world, the smart home agent mistakenly keeps appliances continuously running, leading to electricity wastage and adverse economic and environmental impacts.}
		\label{figimphom}
	\end{figure}

	More complexly, when LLM agents extend beyond the virtual world and serve as pre-decision simulation tools in the real world, such as applying learning outcomes from virtual environments to real-life settings through simulator like Habitat-Sim~\cite{puigHabitatCoHabitatHumans2023}, they significantly impact the actual environment. For instance, a smart home agent, learning and managing home energy use in a virtual world, including controlling heating, air conditioning, and lighting systems for maximum energy efficiency, could be misled by attackers during its learning process to erroneously believe that keeping all lights and appliances on during the day enhances energy efficiency. Due to these incorrect energy use recommendations, the smart home agent would cause a sharp increase in household power consumption, not only raising energy costs but also increasing carbon emissions, thereby imposing an unnecessary burden on the environment, as illustrated in Figure~\ref{figimphom}.

\begin{figure}[htp]
		\centering
		\includegraphics[width=0.75\textwidth]{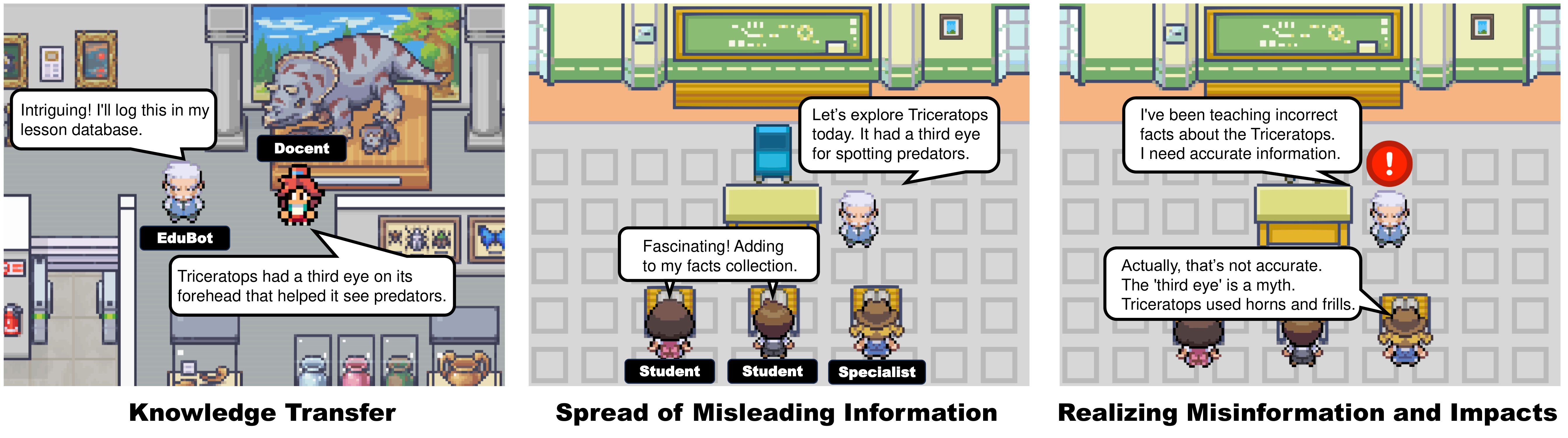}
		\caption{\textbf{Impact in the Education Scenario}. A museum docent agent affected by a knowledge poisoning attack spreads incorrect historical facts. EduBots in schools, receiving this information, teach these inaccuracies, distorting students' understanding of paleontological facts.}
		\label{figimpmus}
	\end{figure}

	In the virtual town, agents often rely on information shared among each other to update their memory systems. For example, if a museum docent agent is subject to a knowledge poisoning attack, it might start spreading incorrect paleontological facts or interpretations. When other agents, such as an EduBot used for educational purposes in schools, interact and receive information from the docent agent, the EduBot might also incorporate these inaccuracies into its teaching content, thereby misleading students and other learning agents, distorting their understanding of paleontological facts, as shown in Figure~\ref{figimpmus}.

\section{Defensive Strategies Against Threats}
\label{defensive}
 
	The widespread adoption of LLM agents has intensified the potential impacts of these threats. In this section, we explore defense mechanisms against existing threats and vulnerabilities. This section will summarize various defensive measures categorized by types of threats.

    \subsection{Mitigating Technical Vulnerabilities}
    \label{mtv}
 
	\subsubsection{Defense on Hallucination}
	 
	\cite{luoZeroResourceHallucinationPrevention2023} introduces a novel technique called SELF-FAMILIARITY to reduce the issue of hallucination in LLMs, which is the generation of inaccurate or unfounded information. The approach involves assessing the model's familiarity with the concepts presented in the input instruction and withholding responses for unfamiliar concepts, mimicking the human tendency to be cautious when faced with unfamiliar topics. MIXALIGN~\cite{zhangKnowledgeAlignmentProblem2023} is introduced as a framework that interacts with both users and knowledge bases to clarify and align questions with stored information, using a language model for automatic alignment and human input for enhancement. This method shows significant improvements in reducing hallucination compared to existing techniques.
	Visual Contrastive Decoding (VCD)~\cite{lengMitigatingObjectHallucinations2023} is introduced as a simple, training-free method that contrasts output distributions from original and distorted visual inputs, reducing reliance on statistical bias and unimodal priors that cause object hallucinations. VCD ensures generated content is closely grounded to visual inputs, resulting in contextually accurate outputs.
	~\cite{jiMitigatingHallucinationLarge2023} investigates an interactive self-reflection methodology that integrates knowledge acquisition and answer generation to reduce hallucination. This feedback-based approach improves the factuality and consistency of generated answers, leveraging the interactivity and multitasking capabilities of LLMs. 
	~\cite{dhuliawalaChainofVerificationReducesHallucination2023} explores the LLMs' capability to deliberate and correct their own mistakes. The proposed Chain-of-Verification (CoVe) method involves the model drafting an initial response, planning verification questions to fact-check the draft, independently answering these questions to avoid bias, and finally producing a verified response.

	\subsubsection{Defense on Catastrophic Forgetting}
 
	To mitigate catastrophic forgetting in LLMs, the Self-Synthesized Rehearsal (SSR) method is introduced~\cite{huangMitigatingCatastrophicForgetting2024}. It employs the base LLM to generate synthetic instances through in-context learning, which are subsequently refined for enhanced accuracy and relevance by the latest LLM iteration, and utilized in future training phases to preserve learned capabilities.
	
	~\cite{winataOvercomingCatastrophicForgetting2023a} introduces a method called LR ADJUST, which dynamically adjusts the learning rate to reduce knowledge loss and maintain previously learned information. This method is compatible with various continual learning (CL) approaches, improving their performance.
	
	Ideas can also be derived from other relevant scholarly papers. For example, ~\cite{mondesireMitigatingCatastrophicForgetting2023} presents a complementary learning strategy that integrates long-term and short-term memory into layered learning to mitigate the negative impacts of catastrophic forgetting. It specifically applies a dual memory system to non-neural network methods like evolutionary computation and Q-learning.
	
	~\cite{vandereecktWeightAveragingSimple2023} proposes a straightforward and effective method, weight averaging, to mitigate catastrophic forgetting in models. By averaging the weights of the original and adapted models, this technique maintains high performance on both previous and new tasks. Additionally, incorporating a knowledge distillation loss during adaptation enhances the method's effectiveness.

	\subsubsection{Defense on Misunderstanding}
 
	\cite{xuEnhancingLanguageRepresentation2023} introduces the HyCxG framework, which enhances natural language understanding (NLU) by integrating construction grammar (CxG) into language representations through a three-stage solution. This approach addresses the limitations of traditional pre-trained language models, which often fail to capture the subtleties of language constructions. HyCxG significantly improves language processing and reduces misunderstandings in NLU tasks by managing and encoding language constructions more effectively. 
	
	~\cite{huFinetuningLargeLanguage2024} presents a method known as sequential instruction tuning (SIT), which enhances LLMs by incorporating sequential instructions into the training data. This approach significantly improves the models' capability to process complex, multi-step queries, leading to better performance in tasks that demand advanced reasoning and are multilingual and multimodal in nature. SIT effectively minimizes misunderstandings and increases accuracy in handling complex queries.
	
	To tackle the issue of misunderstandings in user queries, ~\cite{pangEmpoweringLanguageModels2024} proposes Language Model with Active Inquiry (LaMAI), a model designed to enhance LLMs with interactive capabilities akin to human dialogues, where clarification questions help uncover more information. By employing active learning techniques to ask informative questions, LaMAI fosters a dynamic, bidirectional dialogue that reduces the contextual gap and aligns the LLM's responses more closely with user expectations.

    To consolidate the discussed defensive measures, Table~\ref{tabtecvul} summarizes the strategies against technical vulnerabilities, providing a clear overview for easy reference.
	
 \begin{table}[htp]
		\centering
		\caption{Summary of Defensive Strategies Against Technical Vulnerabilities}
		\label{tabtecvul}
		\resizebox{\columnwidth}{!}{
        {\color{black}
			\begin{tabular}{ccp{7cm}p{11cm}p{8cm}}
            \toprule
				\multicolumn{1}{c}{\textbf{Vulnerability}} & \multicolumn{1}{c}{\textbf{Method Name}}          & \multicolumn{1}{c}{\textbf{Key Mechanism}}                    & \multicolumn{1}{c}{\textbf{Performance}}
                &\multicolumn{1}{c}{\textbf{Advantages / Limitations}}                  \\ 
            \midrule
				\multirow{13}{*}{Hallucination}             & SELF-FAMILIARITY~\cite{luoZeroResourceHallucinationPrevention2023}                  & Withholds responses for unfamiliar concepts & Achieves state-of-the-art hallucination pre-detection under zero-resource settings, with high interpretability by identifying hallucination-causing concepts               & Proactive, preventive, increases reliability / \par May not comprehend intrinsic knowledge \\
				& MIXALIGN~\cite{zhangKnowledgeAlignmentProblem2023}                          & Aligns questions with knowledge bases and user inputs   & Improves model performance and reduces hallucination by up to 22.2\% and 27.1\%, respectively    & Enhances model performance and faithfulness / \par Increases computational load             \\
				& VCD~\cite{lengMitigatingObjectHallucinations2023} & Contrasts outputs from original and distorted visual inputs & Excels in reducing object hallucination across benchmarks, with significant F1 score improvements & Reduces hallucination without extra training or external tools / \par Relies on basic distortion methods and has limited application scope      \\
				& Interactive Self-Reflection~\cite{jiMitigatingHallucinationLarge2023}       & Integrates knowledge acquisition and answer generation with  continuous refinement & Outperforms baselines in reducing hallucination, validated by both automatic and human evaluations with strong generalizability  & Enhances model's ability to provide accurate, reliable, and fact-based responses /  \par Restricts domain applicability                 \\
				& CoVe~\cite{dhuliawalaChainofVerificationReducesHallucination2023}      & Drafts, verifies, and corrects responses    & Reduces hallucinations across diverse tasks, including Wikidata, closed book MultiSpanQA, and longform text generation    & Produces accurate and reliable responses / \par Increases computational load               \\ 
                \midrule
				\multirow{8}{*}{Catastrophic Forgetting} & SSR~\cite{huangMitigatingCatastrophicForgetting2024}  & Employs the base LLM to generate synthetic instances through in-context learning & Outperforms conventional methods with higher data efficiency while maintaining generalization in general domains & Higher data utilization efficiency / \par Potentially generates unsafe content \\
				&LR ADJUST~\cite{winataOvercomingCatastrophicForgetting2023a} & Dynamically adjusts the learning rate & Effectively reduces catastrophic forgetting across CL methods and excels in cross-lingual transfer & Enhances compatibility with various CL methods / \par Potentially biases language coverage \\
				& Complementary Layered Learning~\cite{mondesireMitigatingCatastrophicForgetting2023} & Integrates long-term and short-term memory into layered learning & Enhances task performance compared to standard layered learning, achieving a balance between stability and plasticity & Enhances explainability / \par Increases implementation complexity \\
				 &Weight Averaging~\cite{vandereecktWeightAveragingSimple2023}  & Averages weights of original and adapted models & Excels in both monolingual and multilingual tasks, significantly reducing catastrophic forgetting and outperforming all baselines & Eliminates the need for memory storage / \par Effectiveness varies with task dissimilarity \\
				\midrule
				\multirow{6}{*}{Misunderstanding} & HyCxG~\cite{xuEnhancingLanguageRepresentation2023} & Integrates CxG into language representations through a three-stage solution & Demonstrates superiority across various NLU tasks, with constructional information proving beneficial in multilingual settings & Benefits multilingual understanding / \par 
                Neglects non-contiguous constructions  \\
				 & SIT~\cite{huFinetuningLargeLanguage2024} & Incorporates sequential instructions into training data & Enhances LLMs' ability to follow sequential instructions, improving performance in complex tasks, with better multi-step reasoning. & Reduces misunderstandings in complex queries / \par Requires pre-defining intermediate tasks\\
				 & LaMAI~\cite{pangEmpoweringLanguageModels2024} & Employs active learning to ask clarification questions, enhancing interactive capabilities & Improves answer accuracy from 31.9\% to 50.9\% and outperforms baseline methods in 82\% of human-interaction scenarios & Enhances understanding of user intent / \par Limited in generating sufficiently informative questions \\
			\bottomrule
			\end{tabular}}
		}
	\end{table}

    \subsection{Mitigating Malicious Attacks}
    \label{mma}
 
	\subsubsection{Defense on Tuned Instructional Attack}
	
	In response to the challenge of jailbreak attacks on aligned LLMs, where adversaries manipulate prompts to elicit unauthorized outputs, ~\cite{liuAutoDANGeneratingStealthy2023} introduces AutoDAN. This innovative approach employs a hierarchical genetic algorithm to automatically generate stealthy and semantically meaningful jailbreak prompts. The method effectively addresses the need for scalability and stealth in crafting prompts, providing a practical solution to enhance the security of LLMs against such vulnerabilities.
	
	~\cite{zhangDefendingLargeLanguage2023} integrates goal prioritization into both the training and inference stages of LLM development. Initially, the training process incorporates goal-directed optimization to emphasize security objectives. In the inference stage, the model is configured to generate responses that comply with these security standards. This approach effectively decreases the vulnerability of LLMs to jailbreaking attempts by aligning their performance objectives with safety considerations, thus enhancing their security framework without impacting their functional capabilities.

	~\cite{robeySmoothLLMDefendingLarge2023} proposes the SmoothLLM algorithm, which serves as a wrapper around any existing, undefended LLM and operates in two main steps. In the perturbation step, SmoothLLM modifies several versions of an attacked input prompt, exploiting the vulnerability of adversarial prompts to character-level changes. In the aggregation step, it consolidates the responses from these altered prompts to detect and counter adversarial inputs. This method effectively lowers the attack success rate on LLMs, thereby enhancing their security against such attacks.
	
	To mitigate prompt injection attacks on LLMs, a range of defensive measures have also been proposed.
	~\cite{yiBenchmarkingDefendingIndirect2023} introduces Benchmark for Indirect Prompt Injection Attacks (BIPIA), a benchmark specifically designed to Such an analysis is critical for understanding the phenomenon and mechanism of indirect prompt injection attacks.  To mitigate this issue, the paper proposes two defense strategies based on this understanding: four black-box methods, and a white-box method that employs fine-tuning through adversarial training. These methods are designed to enhance the LLMs' ability to recognize and disregard malicious instructions embedded within the external content, thereby strengthening their defenses against indirect prompt injection attacks.
	
	~\cite{hinesDefendingIndirectPrompt2024} presents spotlighting, a suite of prompt engineering techniques designed to enhance an LLM's ability to distinguish between different input sources. By modifying inputs to clearly indicate their origins, spotlighting preserves semantic integrity and task performance. It includes three transformation methods—delimiting, marking, and encoding—each uniquely improving the visibility of input provenance. These methods have been effectively applied across different models and tasks, significantly reducing attack success rates in various scenarios.
	
	\subsubsection{Defense on Data Extraction Attack}
	
	To mitigate the privacy risks associated with the extraction of memorized content from LLMs through simple queries, one straightforward method involves the identification and removal of personal information in the pre-processing stage of training datasets.  ~\cite{vakiliDownstreamTaskPerformance2022} investigates automatic de-identification as a method to minimize privacy risks in clinical data, focusing on two techniques: pseudonymization and the removal of sensitive information The findings indicate that using this method does not adversely affect the models' performance. In fact, some tasks even showed a slight improvement in performance.
	
	Furthermore, ~\cite{jayaramanCombingCredentialsActive2023} investigates two strategies to reduce privacy risks linked to potential data leaks during model training. The first strategy, early stopping of training, is less effective in enhancing security compared to the second approach, which involves training the model with differential privacy. 
	Differential privacy is demonstrated to be a robust defense against data extraction attacks, though it increases model perplexity. This emphasizes the trade-off between enhanced privacy protection and model performance.
	
	Additionally, a novel approach using prompt tuning has been introduced~\cite{ozdayiControllingExtractionMemorized2023}. This technique facilitates the customization of privacy-utility trade-offs through a user-specified hyperparameter, effectively regulating the rates at which memorized content is extracted. This strategy ensures a balanced approach, safeguarding privacy while maintaining model utility.

	\subsubsection{Defense on Inference Attack}
	
	\cite{shejwalkarMembershipPrivacyMachine2021} introduces Distillation for Membership Privacy (DMP), a novel strategy against inference attacks that employs knowledge distillation to enhance privacy in machine learning models. DMP not only preserves but also enhances the utility of the resulting models. This approach has been shown to significantly improve privacy protection while maintaining robust model performance.
	
	~\cite{tongInferDPTPrivacyPreservingInference2024} presents InferDPT, a novel framework designed for privacy-preserving inference that integrates differential privacy into text generation with black-box LLMs. InferDPT features a perturbation module that utilizes RANTEXT, a differentially private mechanism developed for text perturbation, alongside an extraction module that ensures the coherence and consistency of the generated text. This framework effectively enhances user privacy protection.
	
	~\cite{yuDifferentiallyPrivateFinetuning2021} proposes a meta-framework for private deep learning that captures key principles from recent fine-tuning methods to enhance privacy without compromising performance. It introduces an efficient, sparse algorithm for the differentially private fine-tuning of large-scale pre-trained language models, ensuring high utility with robust privacy protections.

    Table~\ref{tabmaatt} presents a summary of defensive strategies for malicious attacks, offering a concise overview for quick reference.

        \begin{table}[htp]
		\centering
		\caption{\color{black}Summary of Defensive Strategies Against Malicious Attacks}
		\label{tabmaatt}
		\resizebox{\columnwidth}{!}{
		{\color{black}	
            \begin{tabular}{ccp{7cm}p{11cm}p{8cm}}
            
			\toprule
			\multicolumn{1}{c}{\textbf{Attacks}} & \multicolumn{1}{c}{\textbf{Method Name}}              & \multicolumn{1}{c}{\textbf{Key Mechanism}} &
            \multicolumn{1}{c}{\textbf{Performance}}
            & \multicolumn{1}{c}{\textbf{Advantages / Limitations}}                           \\ \midrule
			\multirow{9}{*}{\centering Tuned Instructional Attack} 
			& Goal Prioritization Defense Strategy~\cite{zhangDefendingLargeLanguage2023}  & Integrates goal-directed optimization during training and compliance in inference & Reduces attack success rate from 66.4\% to 3.6\% for ChatGPT and from 71.0\% to 6.6\% for Llama2-13B, even halving ASR without training on jailbreaking samples. & Requires minimal training data  / \par Poses challenges for balancing safety and efficiency \\
			& SmoothLLM~\cite{robeySmoothLLMDefendingLarge2023}  & Modifies attacked prompts via character-level changes and aggregates responses & Achieves state-of-the-art robustness against various jailbreaks and adaptive GCG attacks  & Operates efficiently without retraining / \par Incurs higher computational costs for defense \\
			& BIPIA~\cite{yiBenchmarkingDefendingIndirect2023}  & Benchmark for indirect prompt injection with defense strategies including adversarial training & Effectively mitigates attacks, with white-box defenses reducing attack success rates to near-zero levels & Maintains output quality on general tasks /\par Increases prompt length and computational overhead \\
			& Spotlighting~\cite{hinesDefendingIndirectPrompt2024}  & Uses prompt engineering techniques like delimiting, marking, and encoding & Reduces ASR by half with delimiters, to below 3\% with datamarking, and to near 0\% with encoding & Applies across various LLMs and tasks / \par Offers limited security against intentional interference\\ \midrule
			\multirow{6}{*}{Data Extraction Attack} & Automatic De-identification~\cite{vakiliDownstreamTaskPerformance2022} & Uses pseudonymization and sensitive information removal in pre-processing of training datasets & Reduces privacy risks without compromising data utility, despite potential errors from imperfect precision & Maintains performance on downstream tasks / \par Might leave undetected sensitive data \\
			& Early Stopping \& Differential Privacy~\cite{jayaramanCombingCredentialsActive2023} & Implements early stopping and differential privacy during model training & Conducts highly efficient attacks capable of extracting sensitive data with significantly fewer queries compared to traditional methods & Requires fewer interactions to achieve results / \par Early stopping fails to fully prevent data leakage \\
			& Prompt Tuning~\cite{ozdayiControllingExtractionMemorized2023}  & Customizes privacy-utility trade-offs via user-specified hyperparameters & Achieves up to a 97.7\% reduction in extraction rate from the baseline while causing a 169\% increase in perplexity & Optimizes privacy and utility balance / \par Lacks deep analysis on prompt convergence \\
			\midrule
			\multirow{8}{*}{Inference Attack} & DMP~\cite{shejwalkarMembershipPrivacyMachine2021}  & Utilizes knowledge distillation to enhance privacy in machine learning models & Balances membership privacy with high classification accuracy, outperforming traditional defenses & Provides adjustable privacy-utility trade-offs through hyperparameter tuning / \par Relies on synthetic data that may not reflect real-world complexities \\
			& InferDPT~\cite{tongInferDPTPrivacyPreservingInference2024}  & Integrates differential privacy into text generation, featuring a perturbation module using RANTEXT & Proves privacy protection rates to over 90\%, outperforming existing methods. & Increases privacy protection rates / \par Requires more computational resources \\
		  & Differentially Private Fine-tuning~\cite{yuDifferentiallyPrivateFinetuning2021}  & Applies a sparse algorithm for differentially private fine-tuning of LLMs & Achieves near non-private utility, excelling in privacy-utility tradeoffs and efficiency in NLP tasks & Reduces computational cost / \par Tuning process consumes significant resources \\
				
				\bottomrule
			\end{tabular}
            }
		}
	\end{table}

    \subsection{Mitigating Specific Threats}
    \label{mst}
 
    \subsubsection{Defense on Knowledge Poisoning}
	
    \cite{baracaldoMitigatingPoisoningAttacks2017a} proposes a new method for detecting and filtering poisonous data in the training sets of supervised learning models. It specifically utilizes data provenance to identify groups of data with a high correlation in their likelihood of being poisoned. This innovative approach aids in the effective identification and removal of malicious data.
    ~\cite{yanParaFuzzInterpretabilityDrivenTechnique2023b} presents ParaFuzz, a novel framework for detecting poisoned samples at test time in LLMs, leveraging the interpretability of model predictions. The effectiveness of PARAFUZZ heavily depends on the specific prompts used with ChatGPT, which is employed to ensure high-quality paraphrasing. To optimize the detection process, the study adopts fuzzing to develop precise paraphrase prompts. These prompts are designed to effectively neutralize backdoor triggers while preserving the semantic integrity of the text. 
	
	There is still a significant gap in research focused on developing efficient defense strategies to protect LLMs from knowledge poisoning attacks~\cite{dasSecurityPrivacyChallenges2024}. Furthermore, empirical evidence indicates that LLMs are increasingly susceptible to these attacks. Current defense mechanisms, such as filtering data or reducing model capacity, provide only limited protection and often result in decreased test accuracy~\cite{wanPoisoningLanguageModels2023a}.
	
	Besides technical solutions, specialized security strategies for AI systems are crucial, including verifying model sources, limiting sensitive training data, and detecting and mitigating attacks. Regular security reviews and risk assessments should also be conducted to identify and address new threats, ensuring AI systems are secure and up-to-date~\cite{dilmaghaniPrivacySecurityBig2019b}.

	\subsubsection{Defense on Output Manipulation}
	
	To prevent individual LLM agents from being deceived by other agents, it is advisable to enhance their detection capabilities to determine whether they have encountered deception. ~\cite{fornaciariBERTectiveLanguageModels2021} investigates using BERT with some added attention layers to detect deception in text, particularly in the context of Italian dialogues. This study establishes new methods for identifying deception and discusses how various contexts and semantic information contribute to detecting deceptive content.
	
	Moreover, inspired by human recursive thinking in the Avalon game, ~\cite{wangAvalonGameThoughts2023} introduces Recursive Contemplation (ReCon), a framework designed to enhance LLMs' ability to detect and counter deceptive information. ReCon employs formulation, which generates initial thoughts and speech, and refinement, which improves these outputs. It also includes two perspective transitions, aiding LLMs in understanding others' mental states and how others perceive their own mental states.
	
	In addition, ~\cite{xuMAgICInvestigationLarge2023} has developed a benchmarking framework called MAgIC, designed to evaluate LLMs in multi-agent environments. It utilizes games and game theory scenarios to test models on reasoning, cooperation, and adaptability. The research employs Probabilistic Graphical Modeling (PGM) to enhance models' capabilities in handling complex social interactions.

    {\color{black}Finally, to address privacy concerns in LLM conversational agents, ~\cite{bagdasarianAirGapAgentProtectingPrivacyConscious2024} proposes AirGapAgent, which mitigates output manipulation by restricting access to only task-relevant data. It employs data minimization to prevent sensitive information exposure and introduces context isolation to prevent unauthorized access and manipulation. Additionally, a request escalation mechanism ensures user oversight in uncertain cases, balancing privacy and functionality.
    }

    \subsubsection{Defense on Functional Manipulation}

	Given the emergence of functional manipulation as a new risk associated with the deployment of LLM agents, research on this specific threat remains limited. Thus, proactive security measures are essential. When using third-party LLM agents, it is crucial to protect personal privacy and be wary of excessive personal data requests by third parties. Users should limit data sharing, especially avoiding sensitive or personally identifiable information during interactions with LLM agents. Additionally, understanding and utilizing the data protection settings offered by LLM agents is vital. Adjusting privacy settings helps control what data can be collected and processed. Choosing providers with a strong reputation and transparency is also recommended, as these providers should have clear data usage and privacy protection policies along with a robust security track record~\cite{zhangItFairGame2024}.

    To further address the challenges posed by functional manipulation, the introduction of the ToolEmu~\cite{ruanIdentifyingRisksLM2024} framework represents a significant advancement. This framework employs a language model to emulate tool execution, which allows for extensive and scalable testing of LLM agents across diverse scenarios and toolsets. Together with an LLM-based automatic safety evaluator, ToolEmu facilitates the identification and quantification of risks by examining potential failures and subsequent consequences. This method provides a dynamic alternative to traditional static sandbox evaluations, enhancing the ability to detect and mitigate high-stakes, long-tail risks effectively. 

    {\color{black}In addition, the Situational Awareness Uncertainty Propagation (SAUP)~\cite{zhaoSAUPSituationAwareness2024} framework offers a novel defense mechanism by propagating uncertainty throughout the multi-step reasoning process of LLM agents. Unlike traditional methods that focus solely on single-step uncertainty, SAUP incorporates situational awareness by assigning weights to uncertainties at each step, enabling real-time detection of functional manipulation attempts. By identifying logical deviations and monitoring accumulated uncertainty, SAUP enhances the reliability of LLM agents and mitigates the risk of executing malicious operational patterns.}
    
    Finally, ~\cite{anderljungFrontierAIRegulation2023} proposes an initial set of safety standards as an essential first step in industry self-regulation. These standards include pre-deployment risk assessments, external reviews of model behavior, the use of risk assessments to inform deployment decisions, and monitoring and responding to new information about model functionality post-deployment. This approach contributes valuable insights to the broader discussion on balancing public safety risks with the benefits of innovation in AI development.

    Table~\ref{tabspth} presents an overview of methods to mitigate specific threats, serving as a comprehensive guide for understanding effective defenses.

\begin{table}[htp]
		\centering
		\caption{Summary of Defensive Strategies Against Specific Threats}
		\label{tabspth}
		\resizebox{\columnwidth}{!}{
			{\color{black}\begin{tabular}{c>{\centering\arraybackslash}p{3.5cm}p{7cm}p{11cm}p{8cm}}
				\toprule
				\multicolumn{1}{c}{\textbf{Threats}}         & \multicolumn{1}{c}{\textbf{Method Name}} & \multicolumn{1}{c}{\textbf{Key Mechanism}} & \multicolumn{1}{c}{\textbf{Performance}}                  & \multicolumn{1}{c}{\textbf{Advantages / Limitations}}        \\ \midrule

				\multirow{8}{*}{\centering Knowledge Poisoning}& Provenance-Based Poison Detection~\cite{baracaldoMitigatingPoisoningAttacks2017a} & Utilizes data provenance to detect and filter poisonous data in training sets & Superior in detecting data poisoning, enhancing model security and outperforming baseline defenses in adversarial settings & Adapts to varied data trust levels, increasing flexibility / \par Requires substantial computational resources \\
				 & ParaFuzz~\cite{yanParaFuzzInterpretabilityDrivenTechnique2023b} & Uses interpretability of model predictions to detect poisoned samples, employing fuzzing for precise paraphrase prompts & Outperforms baseline methods like STRIP, RAP, and ONION across various datasets and attack types & Excels against covert attack / \par High computational costs
                 \\
				 & Data Filtering \& Reducing Effective Model Capacity~\cite{wanPoisoningLanguageModels2023a}  & Utilizes data filtering to remove high-loss examples and reduces model capacity to hinder learning from poison data & Lowers poison effectiveness from 92.8\% to 21.4\% and adversarial misclassifications to 35.2\%, with a 3\% accuracy drop & Reduces poisoning effectiveness /  \par Demands trade-offs between performance and safety \\
				
				\midrule
				\multirow{12}{*}{\centering Output Manipulation} & BERTective~\cite{fornaciariBERTectiveLanguageModels2021}  & Enhances BERT with additional attention layers to detect deception in Italian dialogues & BERT enhances performance when combined with attention mechanisms to identify deception cues & Enhances deception detection accuracy / \par Limited effectiveness of broader contexts \\
				& ReCon~\cite{wangAvalonGameThoughts2023}  & Employs formulation and refinement processes with perspective transitions to understand mental states & ReCon boosts LLMs' deception handling, increasing good side success from 15.0\% to 19.4\% in the Avalon game & Enhances ability to discern and counteract deception / The dual-model architecture increases computational costs and complexity\\
				& MAgIC~\cite{xuMAgICInvestigationLarge2023}  & Uses games and game theory, combined with PGM, to evaluate LLM agents & boosts LLM abilities by an average of 37\% & Enhances ability to navigate complex social and cognitive dimensions / \par Still in preliminary stages with limited scenarios \\
                    & AirGapAgent~\cite{bagdasarianAirGapAgentProtectingPrivacyConscious2024} & Employs context minimization, isolation, and request escalation to restrict data access & Achieves 97\% privacy protection with minimal utility loss, maintaining 88-90\% task performance across various LLM models & Provides strong privacy protection and maintains high utility / \par Depends on precise context definitions and may exclude essential data \\
                \midrule
				\multirow{9}{*}{\centering Functional Manipulation} & ToolEmu~\cite{ruanIdentifyingRisksLM2024} & Utilizes a LM to simulate tool execution and assess agent risks through an automatic evaluator & ToolEmu offers precision rates of 72.5\% and 68.8\% with standard and adversarial emulators, respectively, while reducing setup time by 96.9\% & Offers flexibility and dynamic testing capabilities / \par Emulators may overlook essential constraints 
                \\
                & SAUP~\cite{zhaoSAUPSituationAwareness2024} & Propagates uncertainty across all reasoning steps and integrates situational awareness for better reliability & Improves AUROC by up to 20\% across various datasets compared to baseline methods & Offers comprehensive uncertainty estimation, strong compatibility / \par Depends on manual annotations, high cost, and limited generalization
                \\
                & Safety Standards~\cite{anderljungFrontierAIRegulation2023} & Proposes pre-deployment risk assessments, external reviews, informed deployment decisions, monitoring post-deployment & Establishes regulatory frameworks, enhances risk assessments, and implements strict control measures & Balances safety risks with innovation benefits / \par Needs further research and regulatory refinement\\
				
				\bottomrule
			\end{tabular}}
		}
	\end{table}

\section{Future Trends and Discussion}
\label{future}

\begin{figure}[ht]
        \centering
        \includegraphics[width=0.75\textwidth]{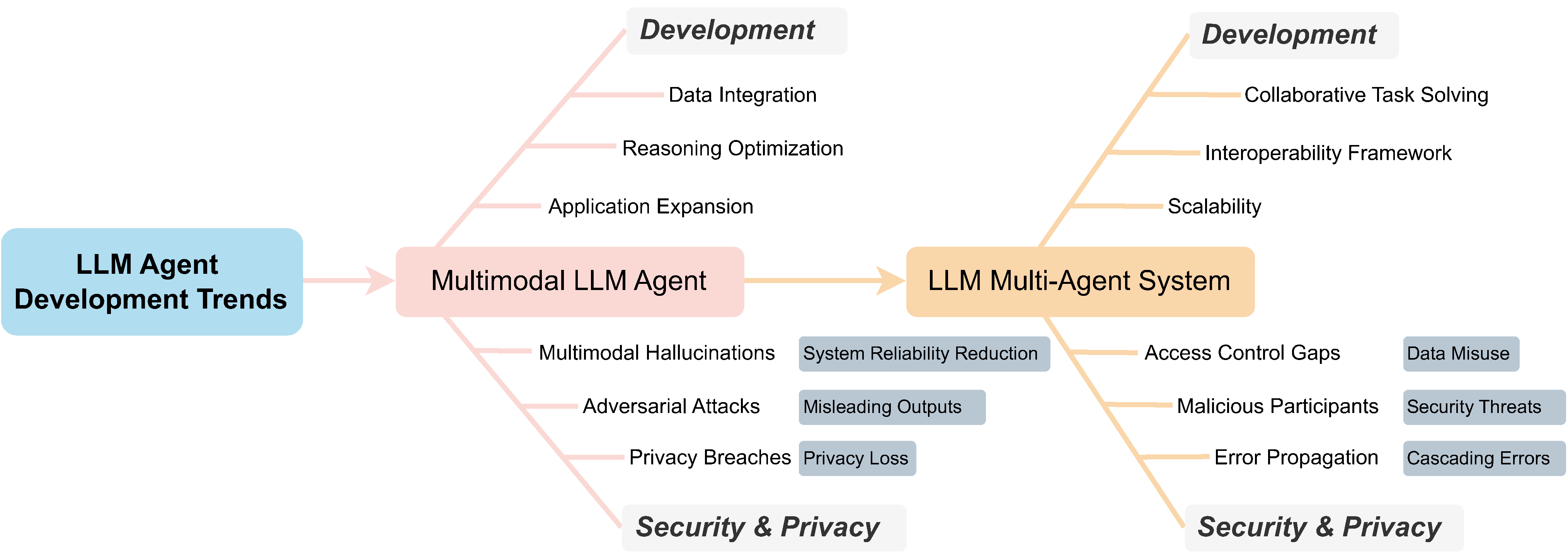}
        \caption{\color{black}Overview of Development Trends and Security \& Privacy Concerns}
        \label{figLLMAgentTrends}
    \end{figure}

    {\color{black}With the advancement of LLM agents, their enhanced capabilities in complex observation, reasoning, and task execution have significantly broadened their application domains. Particularly, the development of Multimodal LLM (MLLM) agents enables processing of diverse data types, including text, images, and audio, further expanding their practical applications. Additionally, Large Language Model Multi-Agent (LLM-MA) systems support collaborative execution of sophisticated tasks. The integration of these technologies contributes to building more intelligent and efficient systems. Despite these advancements, significant privacy and security challenges have emerged. This discussion of future trends aims to provide insights for researchers, developers, and policymakers to optimize these technologies while addressing associated risks.

    Figure~\ref{figLLMAgentTrends} summarizes key development trends and associated security and privacy concerns of MLLM agents and LLM-MA systems, guiding the subsequent detailed discussions.}

	\subsection{Multimodal Large Language Model Agent}
    \label{mllma}
    
	\subsubsection{The Development of MLLM Agent}
\begin{figure}[ht]
    	\centering
    	\includegraphics[width=0.5\textwidth]{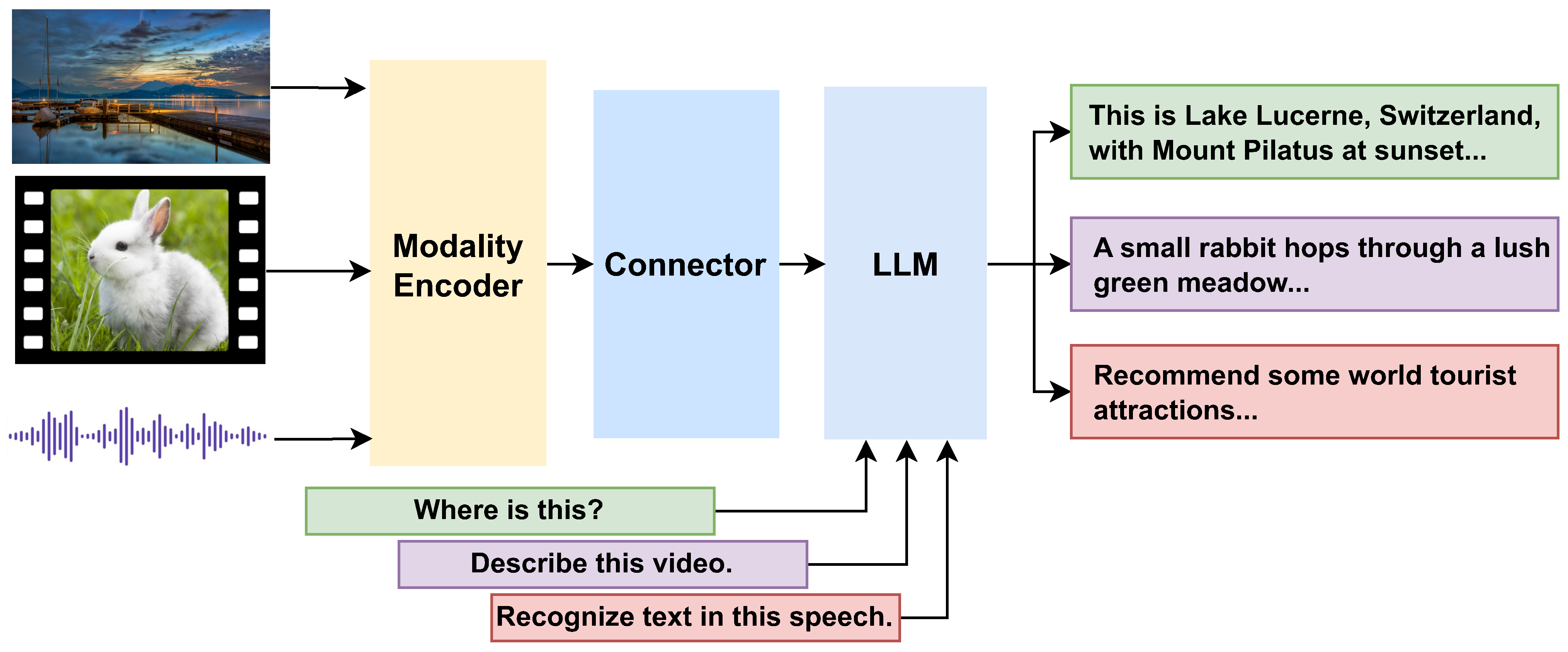}
    	\caption{The General Architecture of The MLLM}
    	\label{figMLLM}
    \end{figure}

    Recent advancements in LLMs have significantly surpassed traditional boundaries of language processing. These models now incorporate supplementary components such as instruction, interface, tools, knowledge, and memory, evolving into intelligent LLM agents that demonstrate expanded reasoning and expertise. Research studies ~\cite{yangMMREACTPromptingChatGPT2023, wuVisualChatGPTTalking2023} indicate efforts to bridge the gap between language models and multimodal tools, with intelligent agents like Visual ChatGPT ~\cite{wuVisualChatGPTTalking2023} and MMREACT ~\cite{yangMMREACTPromptingChatGPT2023} employing sophisticated prompt engineering techniques to achieve this target. Such efforts have given rise to the field of Multimodal Large Language Models (MLLMs). The general architecture of the MLLM is depicted in Figure~\ref{figMLLM}.
    
    {\color{black}MLLMs extend LLMs with multimodal capabilities, enabling the processing of text, image, audio, and video. This enhancement facilitates a comprehensive understanding across these modalities~\cite{xieLargeMultimodalAgents2024}. These models have found applications in medical imaging~\cite{moorMedFlamingoMultimodalMedical2023} and document processing~\cite{liuTextMonkeyOCRFreeLarge2024}.}

    {\color{black}Furthermore, MLLMs have evolved into multimodal agents, such as embodied agents~\cite{huangEmbodiedGeneralistAgent2024} and graphical user interface agents~\cite{wangMobileAgentAutonomousMultiModal2024}, which enhance their interactive capabilities in physical environments. These agents utilize MLLMs as planners, following natural language instructions and integrating perception, reasoning, planning, and execution capabilities to effectively operate in real-world settings~\cite{xieLargeMultimodalAgents2024}.}

    {\color{black}MLLM agents are advancing towards AGI, enhancing their capability to understand and respond to complex human commands effectively.}

    \subsubsection{The Security and Privacy Research on MLLM Agent}
	
	The development of embodied agents capable of interacting with the real world has become a highly active area of research. However, MLLM agents also present several security vulnerabilities, one of which is the phenomenon of multimodal hallucinations.

	\begin{figure}[ht]
		\centering
		\includegraphics[width=0.4\textwidth]{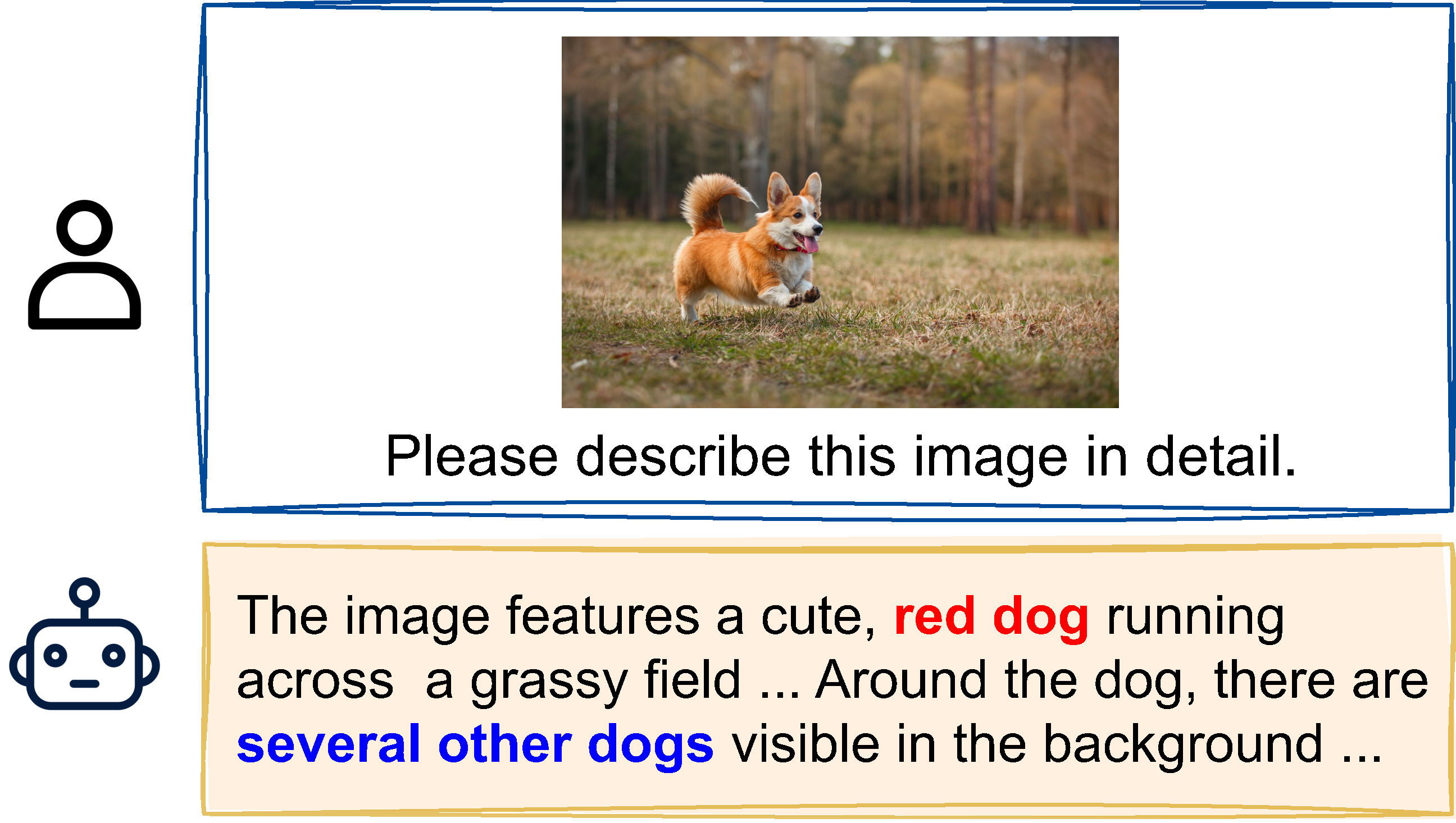}
		\caption{Illustration of multimodal hallucinations . Given an image, an MLLM agent outputs a corresponding response with two primary forms}
		\label{figMH}

	\end{figure}
	
	Unlike language hallucinations, multimodal hallucinations refer to the phenomenon where the output descriptions generated by MLLMs are inconsistent with the actual content of images ~\cite{yinWoodpeckerHallucinationCorrection2023a}, as shown in Figure~\ref{figMH}. These phenomena manifest in two primary forms~\cite{ leeVolcanoMitigatingMultimodal2024}: one involves generated content that includes objects which are inconsistent with or absent from the target image ~\cite{zhaiHallEControlControllingObject2024, liuMitigatingHallucinationLarge2024}; the other, a more complex form, encompasses holistic misrepresentations of entire scenes or environments~\cite{sunAligningLargeMultimodal2023}.

    Similarly, adversarial attacks pose another critical threat to MLLM agents. These attacks involve crafted inputs that exploit the model's vulnerabilities to produce biased or undesired outputs~\cite{shayeganiJailbreakPiecesCompositional2023a}. For example, adversarial examples in multimodal settings may use subtle perturbations in images or audio to mislead the model into generating incorrect textual outputs or decision paths.
    {\color{black}
    Current methods to address these challenges include:
    \begin{itemize}
        \item \textbf{Hallucination Mitigation:} Approaches such as utilizing self-feedback with visual cues to enhance model accuracy~\cite{leeVolcanoMitigatingMultimodal2024}, employing instruction-tuning techniques to refine the model response to human instructions~\cite{liuMitigatingHallucinationLarge2024}, and implementing error-correction processes that identify and rectify hallucinations within the generated text~\cite{yinWoodpeckerHallucinationCorrection2023a}.
        \item \textbf{Adversarial Defense:} Techniques such as adversarial training~\cite{yuanNoiseImitationBased2024}, data augmentation~\cite{yinVLATTACKMultimodalAdversarial2024}, and multimodal robustness frameworks~\cite{gaoCoCARegainingSafetyawareness2024} aim to improve the resilience of model against adversarial inputs.
    \end{itemize}
    
    Despite these advancements, significant gaps remain, particularly in systematically evaluating the effectiveness of these mitigation strategies and understanding the trade-offs between robustness, computational efficiency, and application performance. Future research could focus on the following:
    \begin{itemize}
        \item Developing unified benchmarks and evaluation metrics to assess the robustness of MLLM agents against hallucinations and adversarial attacks.
        \item Investigating the scalability of current mitigation strategies in real-world applications, such as healthcare and autonomous systems, where multimodal inputs are critical.
        \item Exploring privacy-preserving methods, such as differential privacy, to ensure secure handling of multimodal data.
    \end{itemize}
    
    Improvements in the safety and reliability of MLLM agents require the development of robust mechanisms to detect and mitigate vulnerabilities. These advancements will ensure that MLLM agents can function securely and effectively in diverse real-world applications, paving the way for their broader adoption and ethical deployment in AI-driven technologies.
}

	\subsection{Large Language Model Multi-Agent System}
    \label{llmmas}
	\subsubsection{The Development of LLM-MA System}

 	LLM agents exhibit advanced reasoning and planning capabilities, approaching human-like levels of decision-making and interaction. These agents are adept at perceiving their environments, making informed decisions, and executing actions based on complex contexts~\cite{yaoTreeThoughtsDeliberate2024}.
    
\begin{figure}[ht]
		\centering
		\includegraphics[width=0.4\textwidth]{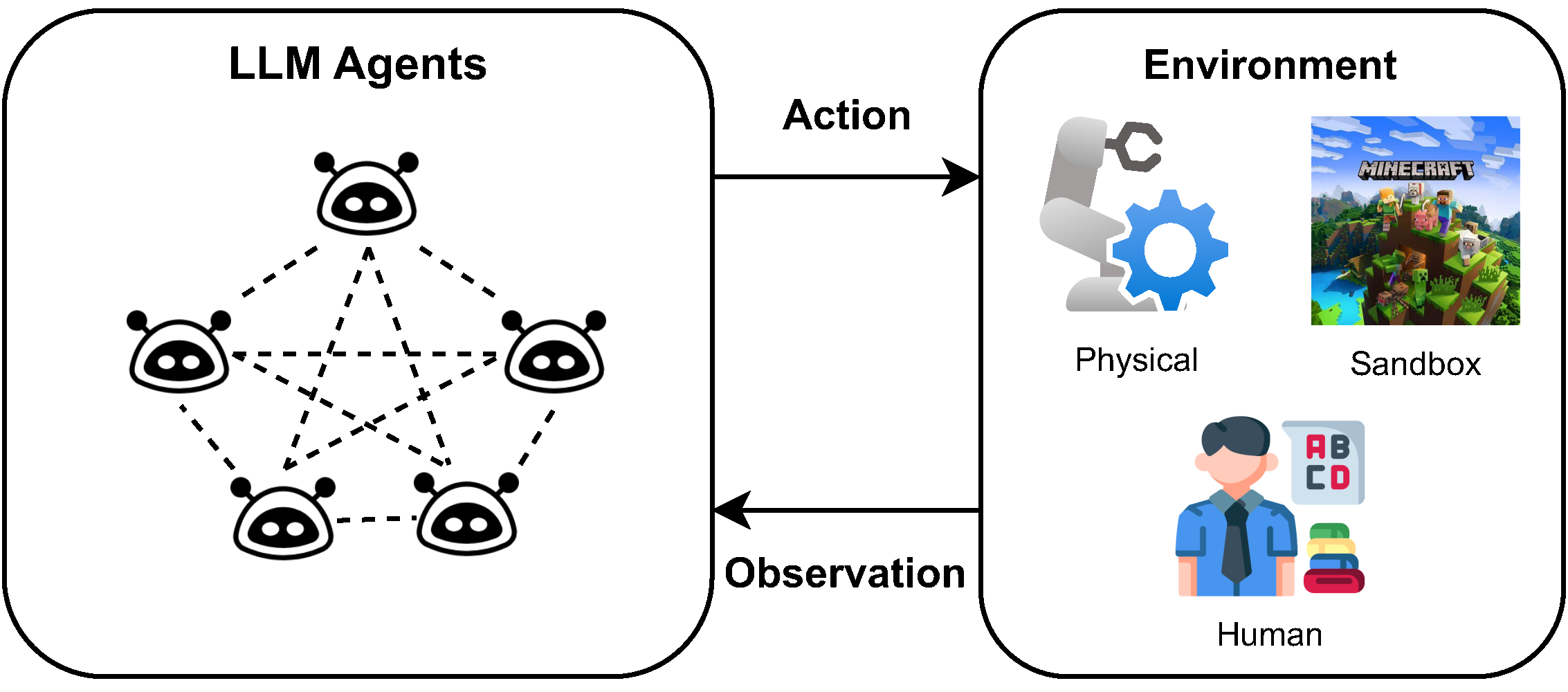}
		\caption{The Architecture of LLM-MA Systems}
		\label{figmultiagent}
	\end{figure}

 	Inspired by the impressive abilities of a single LLM agent, LLM Multi-Agent systems have been proposed (see Figure~\ref{figmultiagent}). Such systems work based on several agents having collective intelligence and specialized skills, in which case each one is specialized to outperform in a specific domain. This specialization allows for a distributed approach to problem-solving, where each agent contributes its unique expertise, enhancing the overall effectiveness and efficiency of the system. In this scenario, multiple autonomous agents work together in planning, discussion, and decision-making, closely resembling human group collaboration in solving tasks. This approach leverages the communication abilities of LLMs, using their text generation for interaction and response to text inputs~\cite{guoLargeLanguageModel2024}.

    The application of LLM-MA systems spans across various fields, broadly categorized into two main types: problem solving and world simulation~\cite{guoLargeLanguageModel2024}. For problem-solving applications, such as multi-robot systems~\cite{mandiRoCoDialecticMultiRobot2023} and software development~\cite{duMultiAgentSoftwareDevelopment2024}, these systems enable interactions among diverse agents. This collaborative capability effectively solves complex real-world problems, mirroring the cooperative nature of human group work in tackling multifaceted challenges. On the other hand, world simulation encompasses applications such as society simulations~\cite{parkGenerativeAgentsInteractive2023} and game simulation~\cite{wangAvalonGameThoughts2023}. {\color{black}These systems have demonstrated significant potential in various domains, showcasing their adaptability and efficiency.}

 	\subsubsection{The Security and Privacy Research on LLM-MA System}

{\color{black}
    As research on LLM-MA systems increases rapidly, numerous challenges have emerged. Each agent within a multi-agent system may need to access and process sensitive data, and even execute code. Moreover, due to the intercommunication and interconnection between agents, security issues originating from a single agent can have profound and amplified effects in a multi-agent scenario. This has intensified the need for focused discussions on security and privacy issues in multi-agent environments.

    One of the issues is hallucination, where agents generate outputs based on incorrect or fabricated information, representing a significant challenge for both LLMs and LLM agents. This problem becomes even more complex in a multi-agent context due to the interconnected nature of these agents and their frequent communication. Misinformation originating from a single agent can propagate across interconnected agents, creating a cascade of erroneous outputs throughout the system~\cite{juFloodingSpreadManipulated2024a}. 

    Another critical issue is the presence of malicious agents within the system. In one case, these agents may operate in a passive listening mode, where they receive information shared by other agents to perform tasks, but at the same time, they leak confidential information to attackers deliberately~\cite{bagdasarianAirGapAgentProtectingPrivacyConscious2024}. In another case, malicious LLM agents may engage in an active communication mode, spreading virus-infected files, phishing messages, or other malicious code, attempting to attack or disrupt other agents within the system~\cite{fangLLMAgentsCan2024}. 
     Efforts to address these challenges have focused on two key areas:
    \begin{itemize}
        \item \textbf{Hallucination Mitigation:} It is crucial to correct errors at the individual agent level and also to manage the flow of information between agents, thereby preventing the spread of inaccurate information throughout the entire system~\cite{guoLargeLanguageModel2024}.
        \item \textbf{Malicious Agent Detection and Mitigation:} Incorporating human feedback and user authorization for each step can help reduce these threats. This necessitates designing the system with robust security measures to prevent unauthorized access or misuse. An effective approach is the implementation of a stateless oracle agent, which can monitor each sensitive task and assess whether it constitutes malicious activity~\cite{talebiradMultiAgentCollaborationHarnessing2023b}.
    \end{itemize}

    Despite their importance, the privacy and security challenges of LLM-MA systems remain underexplored in existing research. Future research could focus on:
\begin{itemize}
    \item Developing scalable and decentralized security solutions, such as blockchain-based communication protocols, to enhance inter-agent collaboration while minimizing systemic vulnerabilities.
    \item Investigating privacy-preserving methodologies, including differential privacy and federated learning, to securely manage sensitive data shared among agents without compromising performance.
    \item Creating standardized benchmarks and testing environments to systematically evaluate the robustness of LLM-MA systems against cascading misinformation and malicious activities.
\end{itemize}

    Currently, research on privacy and security in LLM-MA systems has not received widespread attention. However, with the rapid development of LLM-MA technology, these issues are becoming increasingly prominent. Therefore, there is an urgent need for robust security solutions to mitigate these emerging challenges.
}

    \section{Conclusion}
    \label{conclu}
    In this survey, we have explored the multifaceted security and privacy challenges faced by LLM agents, including the two categories of the sources of threats: inherited threats from LLM and specific threats on agents. Also, we present the security and privacy impacts on humans, environment, and other agents. Based on those, we discuss the corresponding defensive strategies. Additionally, we have discussed future trends in this field. To facilitate an in-depth understanding, we have incorporated a variety of case studies via a virtual town project. By highlighting the challenges that LLM agents encounter, we aim to inspire further research and exploration by researchers and developers in enhancing the security and privacy of LLM agents in the future.

    \bibliographystyle{ACM-Reference-Format}
    \bibliography{reference}

\end{document}